\documentclass[aps,pra,twocolumn,superscriptaddress,10pt]{revtex4-2}

\usepackage[utf8]{inputenc}

\usepackage[dvipsnames]{xcolor} % include additional colours
\usepackage{physics}
\usepackage{soul}
\usepackage{amsmath}
\usepackage{amsfonts}
\usepackage{amssymb}
\usepackage{multirow}
\usepackage{booktabs}
\usepackage{tabularray}
\usepackage[T1]{fontenc}
\usepackage{babel}
\definecolor{mycol}{RGB}{10,55,130}
\usepackage[colorlinks, linkcolor=mycol, citecolor=mycol, urlcolor=mycol, breaklinks]{hyperref}
\usepackage{cleveref}
\usepackage[shortlabels]{enumitem}
\usepackage{empheq}
\usepackage[most]{tcolorbox}
\usepackage{pgffor}
\usepackage{graphicx}
\usepackage[section]{placeins}
\usepackage{bm}
\usepackage{dsfont}
\usepackage[caption=false]{subfig}
\usepackage{bbold}
\usepackage{natbib} %Imports biblatex package
\usepackage{tensor}

\makeatletter
 \@ifundefined{textcolor}{}
 {%
   \definecolor{BLACK}{gray}{0}
   \definecolor{WHITE}{gray}{1} 
   \definecolor{RED}{rgb}{1,0,0}
   \definecolor{GREEN}{rgb}{0,1,0}
   \definecolor{BLUE}{rgb}{0,0,1}
   \definecolor{CYAN}{cmyk}{1,0,0,0}
   \definecolor{MAGENTA}{cmyk}{0,1,0,0}
   \definecolor{YELLOW}{cmyk}{0,0,1,0}
 }

\def\beq{\begin{equation}}
\def\eeq{\end{equation}}

\newtcbox{\mymath}[1][]{%
    nobeforeafter, math upper, tcbox raise base,
    enhanced, colframe=blue!30!black,
    colback=blue!6, boxrule=1pt,
    width=\textwidth,
    #1}

\makeatletter
\def\maketitle{
\@author@finish
\title@column\titleblock@produce
\suppressfloats[t]}
\makeatother

\newcommand{\affA}{QuSoft, Science Park 123, 1098 XG Amsterdam, the Netherlands}
\newcommand{\affB}{Institute for Theoretical Physics, Institute of Physics, University of Amsterdam, Science Park 904, 1098 XH Amsterdam, the Netherlands}

\newcommand{\mytitle}{Open quantum dynamics with variational non-Gaussian states and the truncated Wigner approximation}

\makeatletter
\AtBeginDocument{\let\LS@rot\@undefined}
\makeatother

\begin{document}

\title{\mytitle}

\author{Liam J. Bond}\email{L.J.Bond@uva.nl}\affiliation{\affB}\affiliation{\affA}
\author{Bas Gerritsen}\affiliation{\affB}\affiliation{\affA}
\author{Ji\v{r}\'{i} Min\'{a}\v{r}}\affiliation{\affB}\affiliation{\affA}
\author{Jeremy T. Young}\affiliation{\affB}
\author{Johannes Schachenmayer} \affiliation{CESQ/ISIS (UMR 7006), CNRS and Universit\'{e} de Strasbourg, 67000 }
\author{Arghavan Safavi-Naini}\affiliation{\affB}\affiliation{\affA}

\date{\today}

\begin{abstract}
We present a framework for simulating the open dynamics of spin-boson systems by combining variational non-Gaussian states with a quantum trajectories approach. We apply this method to a generic spin-boson Hamiltonian that has both Tavis-Cummings and Holstein type couplings, and which has broad applications to a variety of quantum simulation platforms, polaritonic physics, and quantum chemistry. Additionally, we discuss how the recently developed truncated Wigner approximation for open quantum systems can be applied to the same Hamiltonian. We benchmark the performance of both methods and identify the regimes where each method is best suited to. Finally we discuss strategies to improve each technique.
\end{abstract}

\maketitle

\tableofcontents

% ===========================================================================================================================
% \tableofcontents
\newpage

% ===========================================================================================================================
% \tableofcontents
\section{Introduction}
Advances in the control of quantum systems over the past decade have led to the development of a wide variety of different platforms for investigating almost-coherent quantum dynamics. However, in the absence of robust fault-tolerant operations, studies of these platforms must contend with various sources of noise induced by couplings to an environment. Moreover, these noisy intermediate-scale quantum (NISQ \cite{Preskill2018}) devices can often be arbitrarily controlled, opening possibilities of creating states far out of equilibrium.
In light of this, understanding the out-of-equilibrium dynamics of open quantum many-body systems has become of great interest, e.g.~for understanding and realizing a quantum advantage in the NISQ era. 

A particularly challenging class of problems arises for
systems containing both spin and bosonic degrees of freedom with an (even locally) unbounded Hilbert space, applicable to quantum simulation and computation platforms ranging from superconducting circuits to trapped ions \cite{Peropadre_2013_PRL, Yoshihara_2017_NatPhys, FornDiaz_2017_NatPhys, Magazzu_2018_NatComm, Marcuzzi_2017_PRL, Gambetta_2020_PRL, Tamura_2020_PRA, Skannrup_2020, Mehaignerie_2023, James_2000_Book, Porras_2004_PRL, Schneider_2012_RepProgPhys, Kienzler_2015_Science, Lo_2015_Nature, Kienzler_2017_PRL}, to paradigmatic problems in impurity physics~\cite{PerezRios2021, LOUS202265}, quantum chemistry~\cite{valahuDirectObservationGeometricphase2023}, or polaritonic chemistry~\cite{garcia2021manipulating}. The ubiquity and complexity of spin-boson Hamiltonians has led to the development of various techniques for their study and characterization. These include methods for the bosonic space, namely path integral techniques \cite{Nalbach_2010_PRB, Kast_2013_PRL, Nalbach_2013_PRB, Otterpohl_2022_PRL}, effective Hamiltonian formulation \cite{Lee_2001_PhysLettB, Rychkov_2015_PRD, SzaszSchagrin_2022_PRD, Rakovszky_2016_NuclPhysB} or lightcone conformal truncation used predominantly in high-energy physics \cite{Anand_2020, Chen_2022_JHEP, Delacretaz_2023_JHEP}, as well as methods such as non-equilibrium Monte Carlo or tensor networks~\cite{del2018tensor, Wellnitz_2022CommPhys, Wall2016}, which have allowed for the simulation of (open) out-of-equilibrium dynamics of quantum many-body systems~\cite{Makri_1995_JChemPhys, Thorwart_1998_ChemPhys, Thorwart_PRE_2000, Schmidt_2008_PRB, Schollwock_2011_AnnPhys, Orus_2014_AnnPhys, Montangero_2018_Book, White_2004_PRL, Schmitteckert_2004_PRB, Nuss_2015_PRB, Dora_2017_PRB, zwolak2004mixed, Daley_2014_AdvPhys, preisser2023comparing}, also in large-system scenarios. However, these methods are often constrained to one-dimensional setups, closed systems, or to a mesoscopic number of particles, or a combination thereof. 

\begin{figure}
    \centering
    \includegraphics[width=0.98\linewidth]{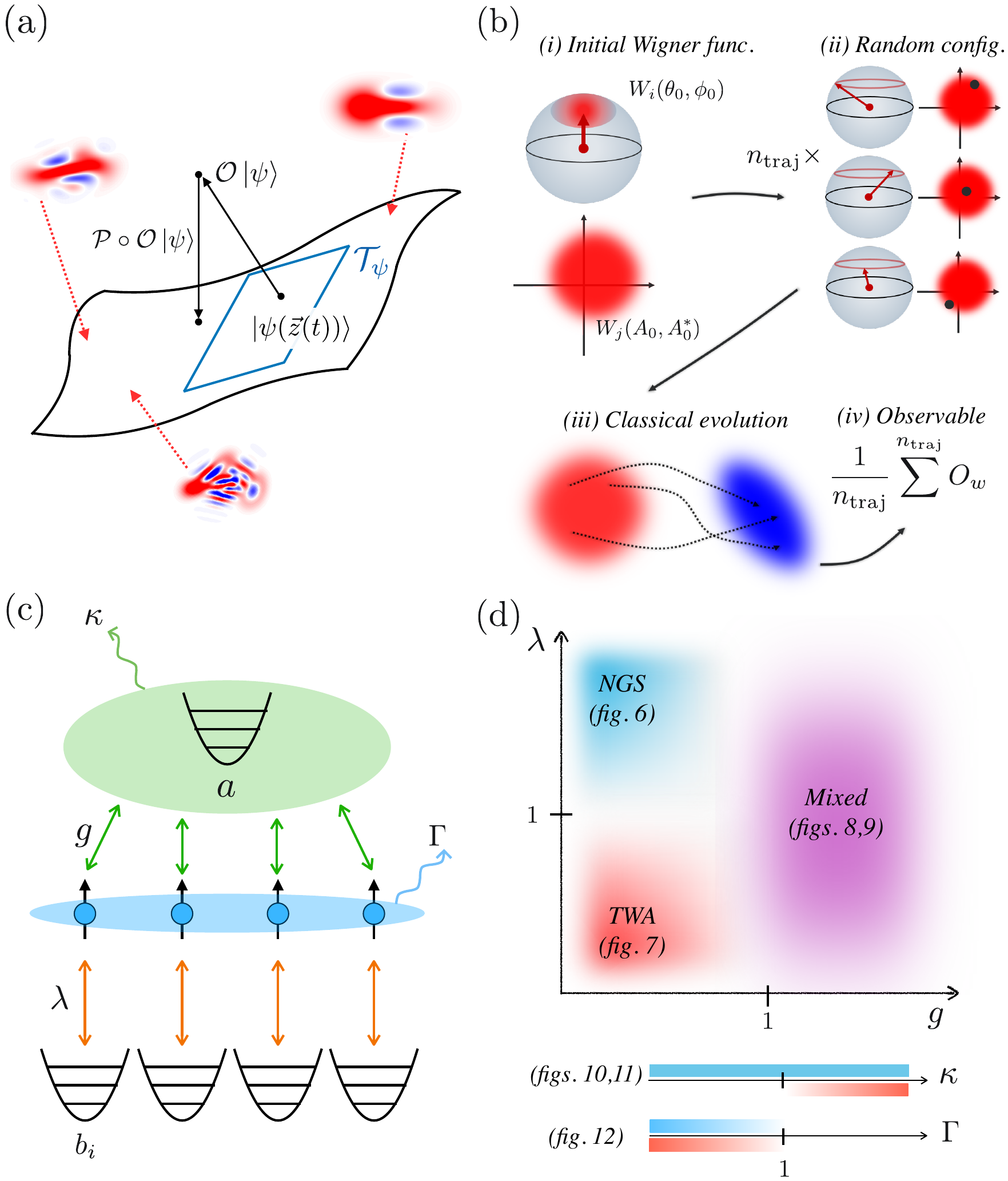}
    \caption{(a) Schematic overview of NGS. The ansatz $\ket{\psi(\vec{z}(t)}$ is a point in a manifold embedded in Hilbert space. For different variational parameters, we can describe a variety of quantum states (some example Wigner functions are shown). The action of an operator $\mathcal{O}$, which can be a Hamiltonian $H$, non-Hermitian Hamiltonian $H_\text{eff}$, or quantum jump $c$, can cause the state to leave the variational manifold, so the state is projected back to the manifold $\mathcal{P}\ket{\psi}$. (b) Schematic depiction of TWA. $(i)$ The initial quantum state of the spin and bosonic degrees of freedom can be represented by the spin Wigner functions $W_i(\theta_0,\phi_0)$ and bosonic Wigner functions $W_j(A_0,A_0^*)$, respectively. $(ii)$  Classical phase space points are individually sampled from the initial Wigner functions for each degree of freedom and $(iii)$ evolve according to the classical (stochastic) equations of motion. $(iv)$ Expectation values of observables are evaluated in phase space by computing the average of the associated Weyl symbols over $n_{\rm traj} $ phase space trajectories. (c) Both NGS and TWA apply to generic spin-boson systems, but we depict here the system studied in this work: $N_s$ spins-$1/2$ interact with a common mode $a$ with strength $g$ and particle loss at rate $\kappa$. Each spin also interacts with a mode $b$ at strength $\lambda$. The spins also undergo collective loss at rate $\Gamma$. (d) Schematic summary of our results illustrating the regions where each method tends to perform well together with the reference to figures studying the dynamics in the respective parameter regimes, see Fig.~\ref{fig:schem} and Sec.~\ref{sec:Results} for more details.}
    \label{fig:methods_schematic}
\end{figure}

The number and breadth of the aforementioned approaches illustrates that no single method can tackle the range of systems described by spin-boson type Hamiltonians or even the full parameter space of a specific model. As such, it is important to pinpoint the strengths and shortcomings of each method. In this work, we undertake a comparative study between two methods: (i) the time-dependent variational ansatz using \emph{non-Gaussian} states (NGS) \cite{Shi_2018_AnnPhys, Hackl_2020_SciPost} and (ii) the truncated Wigner approximation (TWA) combined with its generalization to discrete spaces, discrete truncated Wigner approximation (DTWA). Fig.~\ref{fig:methods_schematic}(a),(b) shows a schematic of each method. These methods may allow us to circumvent some of the aforementioned limitations~\cite{Shi_2018_AnnPhys, Hackl_2020_SciPost, Schachenmayer2015a, Orioli2017,zhu2019a} and have recently been generalized to open quantum systems \cite{Joubert_2015_JChemPhys,Schlegel_2023,Huber2021,Huber2022,Singh2022,FleischhauerDTWA,Mink2023,nagaoSemiclassicalDescriptionsDissipative2024}. Analyzing fundamental quantum effects in macroscopic limits can thus be enabled with NGS and TWA approaches. 

In NGS, one exploits the continuous variable structure of bosonic states to build a time-dependent variational wavefunction ansatz of non-Gaussian states. Here, we specifically use a superposition of squeezed displaced bosonic states, which converges to the true wavefunction due to the over-completeness of the set of coherent states. Since each state in the superposition is Gaussian, much of the previously developed machinery for Gaussian states can be re-utilized. This method has been successfully applied to studies of systems ranging from the Kondo impurity problem \cite{Ashida_2019_PRA}, central spin \cite{Ashida_2019_PRL}, spin-Holstein models \cite{Knorzer_2022_PRL}, Bose and Fermi polarons \cite{Christianen_2022_PRL, Christianen_2022_PRA, Dolgirev_2021_PRX}, and (sub/super) Ohmic spin-boson model~\cite{Bond_2024_PRL}. We also note that the closely related Davydov state ansatz has been applied in the studies of molecular crystals and polaritonic physics \cite{Chen_2023_JChemPhys, Zhao_2023_JChemPhys, Sun_2022_JChemPhys, Zhou_2015_PRB, Zhou_2014_PRB, Wu_2013_JChemPhys,beraGeneralizedMultipolaronExpansion2014}. 

TWA is a semi-classical approach that factorizes the phase space functions (Weyl symbols) that describe a quantum observable in the phase space representation. As such, TWA is reminiscent of a product-state mean-field ansatz on Hilbert space. Like the latter, TWA allows one to treat systems with very large sizes [$\order{10^4}$ particles], while still capturing some essential quantum features such as spin-squeezing or entanglement~\cite{Schachenmayer2015a, lepoutre2019out, Perlin2020, Franke2023, Muleady2023, Young2024}. TWA can be easily adapted to systems with both bosonic and discrete degrees of freedom, combining sampling strategies from continuous~\cite{Polkovnikov2010} and discrete Wigner functions~\cite{Schachenmayer2015a, zhu2019a, Orioli2017}. 

Here, we consider the open and closed dynamics of a spin-boson Hamiltonian featuring multiple spins coupled to a discrete set of bosonic modes via Holstein and Tavis-Cummings couplings, thus ensuring broad applicability of our results. We begin by introducing the two methods: the variational NGS method using the formulation introduced in Ref.~\cite{Bond_2024_PRL} is discussed in Sec.~\ref{sec:NGSclosed}. We extend the method to open quantum systems using the quantum trajectories method in Sec.~\ref{sec:NGStraj}. We discuss TWA with its discrete variant DTWA for closed and open systems in Sec.~\ref{sec:twa}. In Sec.~\ref{sec:Results} we introduce the Holstein-Tavis-Cummings spin-boson Hamiltonian, which we use to compare both methods over a range of parameters. Finally in Sec.~\ref{sec:conclusions} we summarize our findings and discuss how each method can be improved to increase its accuracy and/or applicability both in terms of systems to which they can be applied, and also in terms of the observables that can be accessed. 

% ==============================================================================================================================

\section{Non-Gaussian ansatz for a closed system}\label{sec:NGSclosed}
We begin by introducing the non-Gaussian ansatz, before discussing how to compute the equations of motion for the variational parameters. We consider a system of $N_s$ spins-1/2 and $N_b$ bosonic modes governed by some Hamiltonian $H(t)$. Our wavefunction,  $\ket{\psi(\vec{z})}$, is a variational ansatz in the form of a non-Gaussian state parameterized by a set of real numbers $\vec{z}$,  
\begin{align}
    \ket{\psi(\vec{z})} = \sum_{\sigma=1}^{2^{N_s}}\sum_{p=1}^{N_p} {U}_p^{(\sigma)} \ket{\sigma,0}, 
    \label{eq:psi}
\end{align}
where the summation over $\sigma$ is over all $2^{N_s}$ spin basis states
\footnote{  
The number of spin configurations in the present ansatz scales exponentially with the number of spins and is the bottleneck in the use of the ansatz in Eq.~(\ref{eq:psi}) for many spins. Addressing this issue, such as combining the non-Gaussian ansatz for the bosonic modes with tensor-network techniques for spins, is a matter of future work. We discuss an alternative approach, namely using the Holstein-Primakoff representation of the spins, in Sec.~\ref{sec:TavisCummings}.
}.
The summation over $p$ produces a superposition of $N_p$ bosonic states $U_{p}^{(\sigma)}\ket{0}$ for each spin degree of freedom. We choose the operator $U_{p}^{(\sigma)}$ to be of the form of a \emph{Gaussian} unitary,
\begin{align}
{U}_{p}^{(\sigma)} = e^{\kappa_p^{(\sigma)} + i \theta_p^{(\sigma)}} \mathcal{D}(\vec{\alpha}_p^{(\sigma)}) \mathcal{S}(\vec{\zeta}_p^{(\sigma)}). 
\label{eq:U_Gaussian}
\end{align}
The parameters $\kappa_p$ and $\theta_p$ determine the weight and phase factors, while the many-mode displacement $\mathcal{D}(\vec{\alpha})$ and squeezing $\mathcal{S}(\vec{\zeta})$ operators are defined as, 
\begin{subequations}\begin{align}
\mathcal{D}(\vec{\alpha}) &= \prod_{k=1}^{N_b} \mathcal{D}(\alpha_k) = \prod_{k=1}^{N_b} \exp[\alpha_k a_k^\dagger - \alpha_k^* a_k],  \\ 
\mathcal{S}(\vec{\zeta}) &= \prod_{k=1}^{N_b} \mathcal{S}(\zeta_k) = \prod_{k=1}^{N_b} \exp[\frac{1}{2}\left(\zeta_k^* a_k^2 - \zeta_k {a_k^\dagger}^2\right)], 
\label{eq:S}
\end{align}\end{subequations}
where we dropped the $\sigma$ and $p$ indices of $\vec{\alpha},\vec{\zeta}$ for simplicity. In this article, we always follow the convention that the many-mode operators appear with vector parameters $\vec{\alpha}$ and $\vec{\zeta}$, i.e. $\mathcal{D}(\vec{\alpha})$, $\mathcal{S}(\vec{\zeta})$. Here $a_k$ ($a_k^\dagger$) is the annihilation (creation) operator of the $k$th mode satisfying the commutation relations $[a_j,a_k^\dagger] = \delta_{jk}$. The complex numbers $\alpha_k=x_k + i y_k $ and $\zeta_k= r_k e^{i \phi_k}$ describe the displacement and squeezing amplitudes of the $k$th mode respectively. The parameters $x_k,\, y_k, \, r_k$, and $\phi_k$ are real, and we collect them, along with $\kappa_p$ and $\theta_p$ for each Gaussian state, into a set $\{\kappa,\theta,x,y,r,\phi\}$ indexed by $\ell$. The total set of variational parameters $\vec{z}$ is indexed by four indices: $\sigma \in \{1,\ldots,2^{N_s}\}$, $p \in \{1,\ldots,N_p\}$, $k \in \{1,\ldots,N_b\} $ and $\ell \in \{1,\ldots,6\}$ . The total number of variational parameters is then $M = 2^{N_s} N_p (2+4N_b)$. 

The set of all coherent states forms an over-complete basis for the bosonic Hilbert space. Thus in the limit $N_p \rightarrow \infty$, the NGS ansatz $\ket{\psi}$, even without squeezing ($\vec{\zeta}_p^{(\sigma)} = 0 \; \forall \; p,\sigma$), approaches the true wavefunction $\ket{\Psi}$. The inclusion of diagonal squeezing in our formalism, Eq.~(\ref{eq:U_Gaussian}), is intended to enhance the descriptive power of the ansatz at finite $N_p$. However, in Fig.~\ref{fig:SqVsNoSq} and Sec.~\ref{sec:Results} we show that for the systems studied in this work, a superposition of coherent states (i.e. $\vec{\zeta}$ = 0) for reasonable $N_p$ is typically sufficient to describe the relevant physics. One could further generalize the Gaussian unitary to ${U} \propto \mathcal{D}(\vec{\alpha}) {\rm exp}(-i A^T \mathbb{M} A)$ instead of Eq.~(\ref{eq:U_Gaussian}), where $A=(a_1,\ldots,a_{N_b},a^\dag_1,\ldots,a^\dag_{N_b})^T$ and $\mathbb{M}$ is a symmetric matrix \cite{Shi_2018_AnnPhys, Hackl_2020_SciPost}. However, we find that this significantly complicates all subsequent manipulations, whilst not substantially reducing the $N_p$ required to capture the relevant features of the systems studied in this work. 

We start our analysis by adopting the Dirac-Frenkel variational principle 
\footnote{
A remark is that instead of the Dirac-Frenkel variational principle one could employ the Lagrangian or McLachlan ones. Importantly these coincide, in that they yield the same equations of motion, if the tangent space $\mathcal{T}_{\psi}$ of the variational manifold is a K{\"{a}}hler space}. In this framework, for a given $H(t)$ one can derive equations of motion (EOMs) for the variational parameters $\vec{z}(t)$ describing either real- or imaginary-time evolution of the wavefunction \cite{Hackl_2020_SciPost}, 
\begin{subequations}
\label{eq:EoMs}
\begin{align}
{\rm \text{Real-time} \; ev.:} \;\;\;    \dot{z}^\nu &= -(\omega_{\mu \nu})^{-1} \partial_\mu E(\vec{z},t), \\
{\rm \text{Imag-time} \; ev.:} \;\;\;    \dot{z}^\nu &= -(g_{\mu \nu})^{-1} \partial_\mu \epsilon(\vec{z},t), 
\end{align}
\end{subequations}
where $\mu,\nu = (\sigma, p, k, \ell)$ index the variational parameters $\vec{z}$ and $\partial_\mu = \partial/\partial z^\mu$. Here
$ E(\vec{z},t) = \bra{\psi(\vec{z})}H(t)\ket{\psi(\vec{z})}$ is the energy and $\epsilon(\vec{z},t) = E(\vec{z},t)/\bra{\psi(\vec{z})}\ket{\psi(\vec{z})}
$ is the normalized energy. We introduce the tangent vectors of the variational manifold at point $\vec{z}$, 
\begin{equation}
  \ket{v_\mu} = \partial_\mu \ket{\psi(\vec{z})}.
  \label{eq:v}
\end{equation}
In terms of the tangent vectors, the symplectic form $\omega$ and the metric of the tangent space $g$ are defined as,  
\begin{subequations}
    \begin{align}
    \omega_{\mu \nu} &= 2 \Im \bra{v_\mu}\ket{v_\nu},  \\ 
    g_{\mu \nu} &= 2 \Re\bra{v_\mu}\ket{v_\nu},
    \label{eq:omega}
\end{align}
\end{subequations}
with their inverses denoted $\Omega^{\mu \nu}\equiv\left(\omega_{\mu \nu}\right)^{-1}$ and $G^{\mu \nu}\equiv\left(g_{\mu \nu}\right)^{-1}$. 

Next, we discuss a subtle property of the employed variational principle, following the discussion in Ref.~\cite{Hackl_2020_SciPost}. We assume that the variational parameters are real $\vec{z} \in \mathbb{R}$. Then, the tangent space ${\cal T}_\psi$ of the variational manifold at each point $\ket{\psi(\vec{z})}$ is a \emph{real} vector space spanned by the tangent vectors $\ket{v_\mu}$ embedded in complex Hilbert space. Thus for each basis vector $\ket{v_\mu}$, $i \ket{v_\mu}$ is not guaranteed to lie in the tangent space and has to be projected onto ${\cal T}_\psi$. As the tangent space is not a complex linear space, this projection takes the form 
\begin{align}
    P_\psi i \ket{v_\nu}  = 2 \ket{v_\mu} G^{\mu \sigma} {\rm Re}\bra{v_\sigma} i \ket{v_\nu} = J\indices{^\mu_\nu} \ket{v_\mu},
    \label{eq:projection}
\end{align}
where the complex structure, i.e. the representation of the projection of the imaginary unit, is introduced as $J\indices{^\mu_\nu} = - G^{\mu \sigma} \omega_{\sigma \nu}$. If $J^2 \neq -1$ the projection is non-trivial, that is, $i \ket{v_\mu}$ does not lie in the tangent space. On the other hand, when $J^2 = -1$ on every tangent space, the tangent space of the variational manifold is called a K\"{a}hler space. Then, $i \ket{v_\nu} = J\indices{^\mu_\nu} \ket{v_\mu}$. In this case, $J$ specifies the decomposition of $i \ket{v_\nu}$ on the tangent space vectors $\ket{v_\mu}$. Having a tangent space ${\cal T}_\psi$ to be a \emph{K\"{a}hler space} ensures that the three variational principles (Lagrangian, McLachlan and Dirac-Frenkel) coincide and prevents incorrect couplings in the equations of motion, as shown in Ref.~\cite{Hackl_2020_SciPost}. Finally, we note that satisfying $J^2 = -1$ in the case of real variational parameters is equivalent to requiring a \emph{holomorphic} parametrization in the case of complex parameters. 

In the supplementary material, we prove that $J^2=-1$ for the NGS ansatz, both in the case of a superposition of displaced squeezed states and a superposition of coherent states~\cite{supp}. 

\subsection{Analytic expressions for energies and energy gradients}\label{sec:EnergiesAndEnergyGradients}
Having introduced the NGS ansatz in Eq.~\eqref{eq:psi} and its equations of motion in Eq.~\eqref{eq:EoMs}, we now show how to obtain analytic expressions for the two crucial ingredients to the equations of motion: the energy gradients $\partial_\mu E$ and the geometric structures $g,\omega$. 

Consider a generic spin-boson Hamiltonian. Any such Hamiltonian can be cast in the form $H = \sum (H_\text{s} \otimes H_\text{b})$, where $H_\text{s}$ describes the spin degrees of freedom, and $H_\text{b}$ describes the bosonic degrees of freedom and can be decomposed into monomials of bosonic operators as $H_\text{b} = \bigotimes_{k=1}^{N_b} H_\text{b}^{(k)}$

For a single spin, $N_s=1$, the Pauli matrices $\{\sigma_0,\sigma_x,\sigma_y,\sigma_z\}$ form a complete basis for $H_\text{s}$. For $H_\text{s} = \sigma_z$ and $H_\text{s} = \sigma_x$, we obtain, respectively, 
\begin{subequations}\begin{align}
    \langle \sigma_z H_\text{b} \rangle_\psi &= \sum_{p,p'}^{N_p} \langle {{{U}_{p'}^{(\uparrow)}}}^\dagger H_\text{b} {U}_p^{(\uparrow)} \rangle
    -
    \langle {{{U}_{p'}^{(\downarrow)}}}^\dagger H_\text{b}{U}_p^{(\downarrow)} \rangle, 
    \\ 
    \langle \sigma_x H_\text{b} \rangle_\psi &= \sum_{p,p'}^{N_p} \langle {{{U}_{p'}^{(\uparrow)}}}^\dagger H_\text{b} {U}_p^{(\downarrow)} \rangle
    +
    \langle {{{U}_{p'}^{(\downarrow)}}}^\dagger H_\text{b}{U}_p^{(\uparrow)} \rangle,
\end{align}\end{subequations}
where the expectation values on the right-hand side are evaluated with respect to the bosonic vacuum $\vert 0 \rangle$. Equivalent expressions for $H_\text{s}=\sigma_0$ and $H_\text{s} = \sigma_y$ can be obtained using the same procedure. Thus, the computation reduces to evaluating the many-mode overlap $\langle 0 | {{{U}_{p'}^{(\sigma')}}}^\dagger H_\text{b}{U}_p^{(\sigma)} |0 \rangle$ for each $p,p',\sigma,\sigma'$. Because both the multi-mode displacement $\mathcal{D}(\vec{\alpha})$ and squeezing $\mathcal{S}(\vec{\zeta})$ operators in our ansatz are diagonal in mode operators, we can write the many-mode overlap as a product of single-mode overlaps,
\begin{align}\begin{split}
    &\langle {{{U}_{p'}^{(\sigma')}}}^\dagger H_\text{b}{U}_p^{(\sigma)} \rangle \\ 
    &= \prod_{k=1}^N \bra{0} \mathcal{S}^\dagger(\zeta_{k}^{(p')}) \mathcal{D}^\dagger(\alpha_{k}^{(p')}) H_\text{b}^{(k)} \mathcal{D}(\alpha_{k}^{(p)}) \mathcal{S}(\zeta_{k}^{(p)}) \ket{0},
\end{split}\label{eq:EnergyGradientModeProduct}
\end{align}
where $ H_\text{b}^{(k)}$ is the bosonic operators of $H_b$ that act on mode $k$, as defined above. To evaluate the single mode overlaps in Eq.~(\ref{eq:EnergyGradientModeProduct}) without specifying $ H_\text{b}^{(k)}$, we use the following identity, 
\begin{align}\begin{split}
    \bra{\Theta}{a^{\dagger^{m}}} a^n \ket{\Phi}_s = \left(\frac{\partial}{\partial \beta} \right)^m \left(-\frac{\partial}{\partial \beta^*} \right)^n \chi_g^s(\beta) \bigg|_{\beta=0},
\end{split} \label{eq:poly_a_exp}\end{align}
where $\ket{\Theta},\ket{\Phi}$ are any two quantum states \footnote{note a similar identity holds for open quantum states, obtained by using $\rho$ and the trace} and 
\begin{align}
    \chi_g^s(\beta) = \bra{\Theta}\mathcal{D}(\beta)\ket{\Phi} e^{(s/2) \beta \beta^*},
\end{align}
is  a generalized $s$-ordered characteristic function with $s = 1$ ($s=-1$) denoting (anti-)normal ordering of the bosonic operators $a,a^\dagger$ that appear in the left-hand side of Eq.~(\ref{eq:poly_a_exp}). We note that in the calculation of the single mode overlaps in Eq.~(\ref{eq:EnergyGradientModeProduct}) using Eq.~(\ref{eq:poly_a_exp}), $\ket{\Theta}$ and $\ket{\Phi}$ are single mode Gaussian states. The overlap between any two single mode Gaussian states can be evaluated analytically \cite{PhysRevA.54.5378}, which gives us an analytic expression for $\chi_g^s(\beta)$. Therefore, we can evaluate the partial derivatives in Eq.~\eqref{eq:poly_a_exp} with respect to $\beta,\beta^*$ for any $m,n$ and find analytic expressions for the energy of any Hamiltonian that is polynomial in $a,a^\dagger$ operators. From this, we can also obtain analytic expressions for the energy gradients, $\partial_\mu E$, as needed in the EOMs. 

~\\
\textbf{Example: Energy gradient of harmonic oscillator}
To demonstrate the above machinery we compute the energy gradient of the harmonic oscillator $H = a^\dagger a$ with respect to the NGS ansatz with coherent states only, i.e. $\vec{\zeta} = 0$. We set $N_p = 2$, so $\ket{\psi} = e^{\kappa_1 + i \theta_1}\ket{\alpha_1} + e^{\kappa_2 + i \theta_2}\ket{\alpha_2}$, with $\alpha_i = x_i + i y_i$. The energy is \begin{align*} 
E=\langle H \rangle &= e^{2\kappa_1}(x_1^2 + y_1^2) + e^{2\kappa_2}(x_2^2 + y_2^2) \\
&+ \left(e^{\kappa_1 + \kappa_2 + i (\theta_1 - \theta_2)}(x_1 + i y_1)(x_2 - i y_2) + \text{h.c.}\right),
\end{align*} 
and its partial derivatives with respect to $x_1$ and $\kappa_1$ are given by $\partial_{x_1}E = 2e^{2 \kappa_1 }x_1 + [e^{\kappa_1 + \kappa_2 + i(\theta_1 - \theta_2)}(x_2 - i y_2 ) + \text{h.c.}]$ and $\partial_{\kappa_1} E = 2 e^{2 \kappa_1}(x_1^2 + y_1^2) + [e^{\kappa_1 + \kappa_2 + i(\theta_1 - \theta_2) } (x_1 + i y_1)(x_2 - i y_2)]$, respectively. The partial derivatives $\partial_{y_1} E$, $\partial_{x_2} E$, $\partial_{y_2} E$, $\partial_{\kappa_2}E$, $\partial_{\theta_1}E$, and $\partial_{\theta_2}E$ can be evaluated using the same procedure. We note that these expressions can be extended to any $N_p$, and to any bosonic Hamiltonian that is polynomial in $a,a^\dagger$ using Eq.~(\ref{eq:poly_a_exp}). 

\subsection{Tangent vectors and the overlap matrix}\label{sec:TangentVectorsOverlapMatrix}
Next, we explain how to compute the tangent vectors of the ansatz, $\ket{v_\mu} = \partial_\mu \ket{\psi(\vec{z})} = \partial_{\sigma,p,k,\ell}\ket{\psi(\vec{z})} $. Plugging Eq.~(\ref{eq:psi}) into Eq.~(\ref{eq:v}) we find, 
\begin{align}\begin{split}
    \ket{v_\mu} &=  \frac{\partial}{\partial z_{\sigma,p,k,\ell}} U_p^{(\sigma)} \ket{\sigma,0} \\ 
    &= \prod_{k'\neq k}^{N_b}\biggl[\mathcal{D}(\alpha_{k'})\mathcal{S}(\zeta_{k'})\biggr]\\
    & \hskip10pt \times \frac{\partial}{\partial z_{\sigma,p,k,\ell}}\biggl[e^{\kappa + i \theta}\mathcal{D}(\alpha_{k})\mathcal{S}(\zeta_{k})\biggr]\ket{\sigma,0}.
\end{split}\label{eq:tangentvector}\end{align}
\begin{comment} 
\end{comment} 

The calculation involving many modes and a superposition of squeezed-displaced states now reduces to calculating the tangent vector of a single-mode Gaussian state for a single Gaussian, i.e. $\frac{\partial}{\partial z_{\sigma,p,k,\ell}}e^{\kappa + i \theta}\mathcal{D}(\alpha_{k})\mathcal{S}(\zeta_{k})\ket{0}$. We proceed as follows: (i) using the disentangled forms of the displacement and squeezing operators, we normal order $\mathcal{D}(\alpha_{k})\mathcal{S}(\zeta_{k})\ket{0}$ such that only $a^\dagger$ operators remain, which (ii) enables us to obtain concise analytic expressions for the tangent vectors of $\mathcal{D}(\alpha_{k})\mathcal{S}(\zeta_{k})\ket{0}$, and (iii) obtain expressions for the general NGS ansatz and outline the computation of the overlap matrix $\omega_{\mu \nu}$. 

The disentangled forms of the single-mode displacement and squeezing operators are given by, 
\begin{align}
\mathcal{D}(\alpha) &= e^{\alpha a^\dagger - \alpha^* a} = e^{- \frac{ \abs{\alpha}^2}{2}}e^{\alpha a^\dagger}e^{- \alpha^* a}, \\
\mathcal{S}(\zeta) &= e^{\frac{1}{2} (\zeta {a^\dagger}^2-\zeta^* a^2)} = e^{ \bar{r} {a^\dagger}^2} e^{- {\check{r}} (a^\dagger a + \frac{1}{2})} e^{-\bar{r} a^2},
\end{align}
where $\zeta = r e^{i \phi}$, $\bar{r} = e^{i \phi} \tanh(r)/2$ and $\check{r} = \ln(\cosh(r))$. When applied to $\vert \psi(\vec z)\rangle$, $\mathcal{S}(\zeta)$ always acts directly on the bosonic vacuum which simplifies its action to, 
\begin{align}\begin{split}
\mathcal{S}(\zeta)\ket{0} &= \frac{1}{\sqrt{\cosh{r}}} e^{\bar{r} {a^\dagger}^2} \ket{0}.
\end{split}\label{eq:S(zeta)vacuum}\end{align}
After normal-ordering $\mathcal{D}(\alpha)\mathcal{S}(\zeta)\ket{0}$, we arrive at
\begin{align}\begin{split}
    \mathcal{D}(\alpha)\mathcal{S}(\zeta)\ket{0} &= \frac{e^{-\frac{\abs{\alpha}^2}{2}}e^{\bar{r} {\alpha^*}^2}}{\sqrt{\cosh(r)}}  e^{\alpha a^\dagger} e^{\bar{r} ({a^\dagger}^2 - 2 \alpha^* a^\dagger)}
 \ket{0},
 \label{eq:normal_ordered}
\end{split}\end{align}
where we used the relation $e^{x a} f(a,a^\dagger) e^{-x a} = f(a, a^\dagger + x)$ \cite[\S3.3 Theorem 2]{louisell1973quantum} and the fact that $e^{-\alpha^* a} \ket{0} = \ket{0}$. Note that this expression contains only $a^\dagger$ and thus all terms commute, enabling us to take the derivative with respect to the variational parameters to obtain the tangent vectors. \\

~\\
\emph{Computing tangent vectors.}
For each $p$, $\sigma$ and $k$ there are six variational parameters: $\{\kappa,\theta,x,y,r,\phi\}$. The tangent vectors for the norm $\kappa$ and phase $\theta$ are simply
\begin{subequations}\begin{align}
\partial_\kappa U_p^{(\sigma)} \ket{\sigma,0} = U_p^{(\sigma)} \ket{\sigma,0}, \\ 
\partial_\theta U_p^{(\sigma)} \ket{\sigma,0} = i U_p^{(\sigma)} \ket{\sigma,0},
\end{align}\end{subequations}
and are independent of the mode number $k$. We use the normal-ordered form of $\mathcal{D}(\alpha)\mathcal{S}(\zeta)\ket{0}$, Eq.~(\ref{eq:normal_ordered}), and define $\ket{\alpha,\zeta}\equiv \mathcal{D}(\alpha)\mathcal{S}(\zeta)\ket{0}$ to find the tangent vectors for $x$, $y$, $r$ and $\phi$, 
\begin{subequations}\begin{align}
\partial_{x} \ket{\alpha,\zeta} &= (f_{x} + g_{x} a^\dagger ) \ket{\alpha,\zeta}, \\ 
\partial_y \ket{\alpha,\zeta}&= (f_{y} + g_{y}  a^\dagger) \ket{\alpha,\zeta}, \\ 
\partial_r \ket{\alpha,\zeta}&= ( f_{r} + g_{r} a^\dagger + h_{r} {a^\dagger}^2)  \ket{\alpha,\zeta}, \\ 
\partial_\phi \ket{\alpha,\zeta}&= (f_{\phi} + g_{\phi} a^\dagger + h_{\phi} {a^\dagger}^2 ) \ket{\alpha,\zeta},
\end{align}\label{eq:dxUa}\end{subequations}
where we have introduced the following $c$-number functions $f,g$: 
\begin{subequations}\begin{align}
f_{x} &= e^{i \phi} \tanh(r) (x-iy) -x, \\ 
g_{x} &=  1 - e^{i \phi} \tanh(r), \\ 
f_{y} &= -e^{i \phi} \tanh(r)(y+ix) -y, \\ 
g_{y} &= i(1+e^{i \phi}\tanh(r)), \\ 
f_{r} &= (1/2)[-\tanh(r) + e^{i \phi} \sech(r)^2(x-iy)^2], \\ 
g_{r} &= (1/2)e^{i \phi}\sech(r)^2(2iy-2x), \\ 
h_{r} &= (1/2)e^{i \phi} \sech(r)^2, \\ 
f_{\phi} &= (i/2)e^{i \phi}\tanh(r)(x-iy)^2, \\ 
g_{\phi} &= (i/2)e^{i \phi}\tanh(r)(2iy-2x), \\
h_{\phi} &= (i/2)e^{i \phi} \tanh(r).
\end{align} \label{eq:fgh}\end{subequations}

Note that further simplification of the terms with creation operators in Eqs.~(\ref{eq:dxUa}) in the form of $\mathcal{D}(\alpha)\mathcal{S}(\zeta) \ket{1}$ and $\mathcal{D}(\alpha)\mathcal{S}(\zeta) \ket{2}$ is not possible as we imposed that $\mathcal{S}(\zeta)$ must act directly on the vacuum to obtain the simplified version of $\mathcal{S}(\zeta)\ket{0}$ in Eq.~(\ref{eq:S(zeta)vacuum}). 

Eqs.~(\ref{eq:dxUa}) are simple analytic expressions for the single mode tangent vectors of $\mathcal{D}(\alpha)\mathcal{S}(\zeta)\ket{0}$. When combined with Eq.~(\ref{eq:tangentvector}), we can therefore construct all the tangent vectors for any $\sigma$, $p$ and $k$.\\

~\\
\emph{Computing tangent vector overlaps to construct geometric structures.}
Having obtained the tangent vectors, we finally briefly comment on the calculation of the overlap matrix $\bra{v_\mu}\ket{v_\nu}$, which is used to construct the metric $g_{\mu \nu}$ and sympletic form $\omega_{\mu \nu}$. We note that due to Eq.~(\ref{eq:tangentvector}), the overlap between two tangent vectors $\bra{v_\mu}\ket{v_\nu}$ can be written as a product of single-mode overlaps. Each single-mode overlap can be evaluated using the expressions in Eq.~(\ref{eq:dxUa}) and applying Eq.~(\ref{eq:poly_a_exp}) to evaluate expectation values of the type $\bra{\alpha',\zeta'} (a^\dag)^m a^n \ket{\alpha, \zeta'}$.
\begin{figure}
    \centering
    \includegraphics[width=\linewidth]{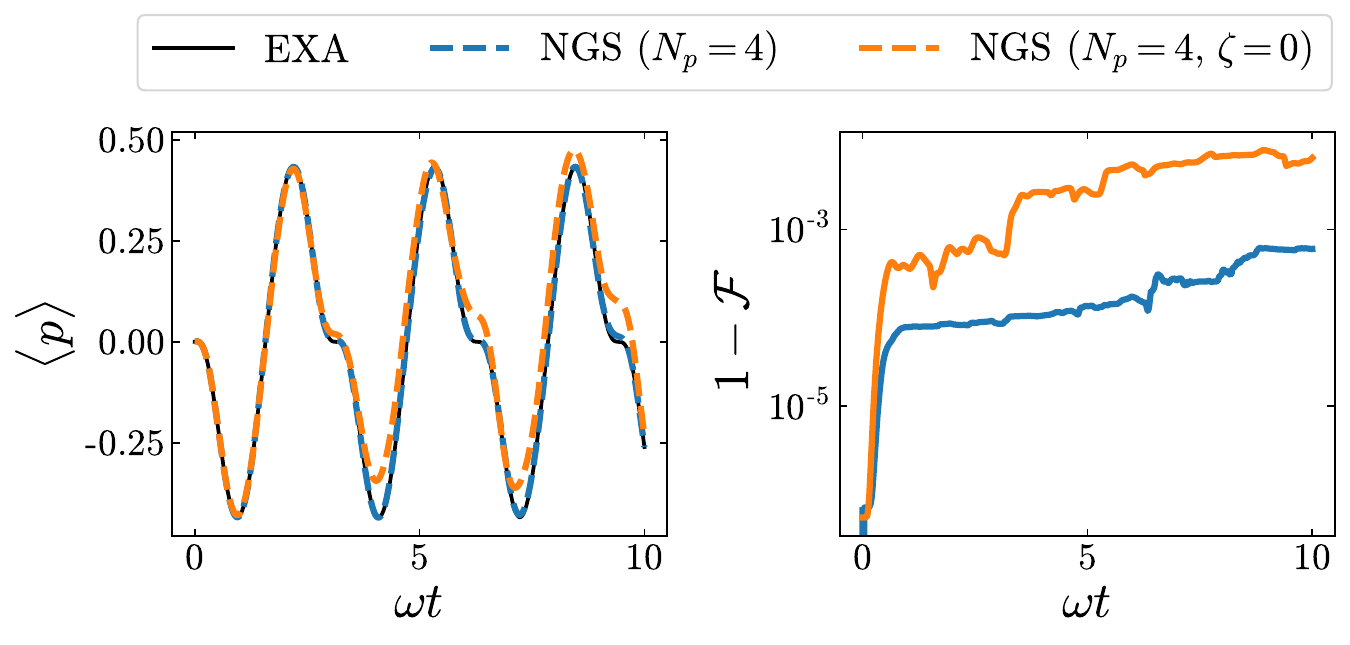}
    \caption{Comparison between the NGS ansatz Eq.~(\ref{eq:psi}) with a superposition of displaced squeezed states vs. a superposition of coherent states, for evolution under the anharmonic oscillator, both with $N_p = 4$. The initial state is $\ket{\psi(0)} = \ket{\alpha=1}$. The inclusion of squeezing (blue dashed) results in closer agreement with exact numerics (solid black) compared to the absence of squeezing in the ansatz (orange dashed). However, both capture the dynamics with small infidelities $1-\mathcal{F} < \mathcal{O}(10^{-2})$. Here ${\cal F}(t)=|\langle \Psi(t) | \psi(t) \rangle|$, where $\ket{\Psi(t)}$ is the quantum state obtained with exact numerics.
    }
    \label{fig:SqVsNoSq}
\end{figure}

~\\
\textbf{Example: Quantum anharmonic oscillator}
As an example of the results of the above machinery, and to illustrate the role of squeezing in the ansatz, we use Eq.~(\ref{eq:EoMs}) and solve for the dynamics of a simple anharmonic oscillator, $H = \omega a^\dagger a + \mu (a^\dagger a)^2$ ~\cite{Milburn_1986_PRA, Halpern_1973_JMathPhys, Bender_1973_PRD, Bender_1969_PR}. Results for the strong coupling regime setting $\omega = \mu = 1$ are shown in Fig.~\ref{fig:SqVsNoSq} with $N_p=4$. We observe that the ansatz containing squeezing (dashed blue line) has a better agreement with the actual state $\ket{\Psi}$ (solid black line, obtained from exact numerical solution of the Schr\"{o}dinger equation) than the ansatz without squeezing (dashed orange line). While this observation always depends on the specific system being studied, we observe that for many applications in this work the improvement in accuracy from including squeezing comes at the expense of additional computational resources. Motivated by this, for the remainder of this article our results utilize the NGS ansatz of a superposition of (many-mode) coherent states.

\section{Open Dynamics: Combining NGS with quantum trajectories}\label{sec:NGStraj}
To extend the NGS machinery to open dynamics there are several options available. For instance, in Ref.~\cite{Schlegel_2023}, which in turn builds on the developments in Ref.~\cite{Joubert_2015_JChemPhys}, an ansatz for the density matrix was developed. This was then used to formulate the master equation and find the corresponding equations of motion governed by the Lindbladian. 

Here, we discuss how the ansatz introduced previously, $\ket{\psi(\vec{z})}$ in Eq.~\eqref{eq:psi}, can be used within the quantum trajectories framework.
In Sec.~\ref{sec:QuantumTrajectories} we recall the basic formulation of the quantum trajectories approach before formulating equations of motion for the effective non-Hermitian Hamiltonian in Sec.~\ref{sec:Open_EoMs}. We give a recipe for how to formulate the action of quantum jumps within the NGS ansatz in Sec.~\ref{sec:QuantumJumps}, and we finally discuss the relevant issues related to the implementation of the equations of motion in Sec.~\ref{sec:Trottetisation}. 
To elucidate the formalism, we provide specific examples throughout this section. Our examples are motivated by typical sources of decoherence in spin-boson quantum simulation platforms governed by Hamiltonians of the form that will be introduced in Sec.~\ref{sec:Results}. 

\subsection{Quantum trajectories}\label{sec:QuantumTrajectories}
Many problems of dynamics of open quantum systems are amenable to the standard time-local master equation in the Lindblad form \cite{Breuer_2002_Book, Weiss_2012_Book, Gardiner_2004_Book}
\begin{equation}
    \dot{\rho} = -i [H,\rho] + \sum_m c_m \rho c_m^\dag - \frac{1}{2} \left\{ c_m^\dag c_m, \rho \right\},
    \label{eq:master}
\end{equation}
where $c_m$ are the jump operators and $\{ \cdot, \cdot\}$ the anticommutator. 

An alternative approach is to use the \emph{quantum trajectories} method \cite{Dalibard_1992_PRL, Molmer_1993_JOptB, Plenio_1998_RMP, Daley_2014_AdvPhys}. Here, rather than evolving the full density matrix described by $2^{2 N_s}$ elements (for $N_s$ spin-1/2 particles), one \emph{stochastically} evolves a pure quantum state $\ket{\psi}$ described by $2^{N_s}$ elements. This approach consists of two steps: (i) continuous evolution under the Schr\"{o}dinger equation with an effective \emph{non-Hermitian} Hamiltonian, 
\begin{align}
    H_{\rm eff} = H-\frac i2 \sum_m c_m^\dag c_m=H - i K.
    \label{eq:Heff2}
\end{align}
and (ii) discrete evolution under the action of a quantum jump operator $c_m$.  Such stochastic evolution of the wavefunction $\ket{\psi}$ constitutes a quantum trajectory. Observables of interest are then evaluated by averaging over $n_{\rm traj}$ such quantum trajectories. In scenarios where it is possible to obtain the observables of interest with sufficient accuracy using $n_{\rm traj} < 2^{N_s}$, quantum trajectories can be more efficient as compared to solving the master equation \cite{Molmer_1993_JOptB, Daley_2014_AdvPhys, preisser2023comparing}.

Our key contribution is to perform both (i) and (ii) using NGS. The details of the quantum trajectories method, and a step by step description of its implementation, can be found in Ref.~\cite{Daley_2014_AdvPhys}. 

\subsection{Equations of motion for non-Hermitian Hamiltonians}
\label{sec:Open_EoMs}
In between the quantum jumps, the wavefunction evolves continuously under the effective non-Hermitian Hamiltonian Eq.~(\ref{eq:Heff2}). Here, we follow the procedure in Ref.~\cite{yuanTheoryVariationalQuantum2019} to derive equations of motion for evolution under $H_{\rm eff}$. Note that the same procedure is used to derive the real-time and imaginary-time equations of motion given in Eqs.~(\ref{eq:EoMs}) describing evolution under a Hermitian Hamiltonian $H$. The NGS wavefunction evolves according to the Schr\"odinger equation, 
\begin{align}
    \partial_t \ket{\psi} = -i H_{\rm eff} \ket{\psi}. 
\end{align}
McLachlan's variational principle requires the variation of the norm resulting from the Schr\"{o}dinger equation to vanish, 
\begin{align}
    \delta || (d/dt + i H_{\rm eff} ) \ket{\psi} || = 0.
    \label{eq:Open_McLachlan}
\end{align}
Since it is more convenient to work with the square norm, we rewrite Eq.~(\ref{eq:Open_McLachlan}) as $\delta || (d/dt + i H_{\rm eff} ) \ket{\psi} ||^2 = 0$, obtaining
\begin{align}
\begin{split}
    & || (d/dt + i H_{\rm eff} ) \ket{\psi} ||^2 = \sum_{\mu,\nu} \bra{v_\mu} \ket{v_\nu} \dot{z}_\mu \dot{z}_\nu \\ 
    &+ \left( i\bra{v_\mu} H_{\rm eff} \ket{\psi} \dot{z}_\mu + \text{h.c.} \right) + \bra{\psi}{H_{\rm eff}}^\dagger{H_{\rm eff}}\ket{\psi}.
    \end{split}
\end{align}
After making the following substitutions 
\begin{subequations}
    \begin{align}
        A_{\mu \nu} &= \bra{v_\mu}\ket{v_\nu}, \\
        C_\mu &= \bra{\psi} H \ket{v_\mu},\label{eq:C}\\
        D_\mu &= \bra{\psi} K \ket{v_\mu},\label{eq:D}
    \end{align}
\end{subequations}
the variation of the square of the norm is
\begin{align}
\begin{split}
    & \delta || (d/dt + i H_{\rm eff} ) \ket{\psi} ||^2
    = \sum_\mu \sum_{\nu} \left(A_{\mu \nu} + \text{h.c.} \right) \dot{z}_\nu \delta z_\mu  \\  
    & + \left[ i (C_\mu^\dagger - i D_\mu^\dagger) + \text{h.c.} \right] \delta z_\mu, 
    \end{split}
\end{align}
which yields the equations of motion,
\begin{align}
    \sum_\nu \Re[A_{\mu \nu} ] \dot{z}_\nu &= \Re[D_\mu] + \Im[C_\mu]. 
    \label{eq:EoMs_Heff}
\end{align}

The object $2 \Re[A_{\mu \nu}]$ is precisely the metric $g$ of the tangent space introduced in Eq.~\eqref{eq:omega}, which we showed how to compute in Sec.~\ref{sec:TangentVectorsOverlapMatrix}. $\Re[D_\mu]$ can be related to $\partial_{\mu} \langle K \rangle$ via
\begin{align}
    2 \Re[ \bra{v_\mu} K \ket{\psi} ] = \bra{v_\mu} K \ket{\psi} + \text{h.c.} = \partial_{\mu} \langle K \rangle,
\end{align}
where we substituted the definition of the tangent vector into the definition of $D$ in Eq.~\eqref{eq:D}, and used the fact that $K = K^\dagger$. We showed how to analytically compute the gradient of expectation value such as $\partial_{\mu} \langle K \rangle$ in Sec.~\ref{sec:EnergiesAndEnergyGradients}. 

Although $\Im[C_\mu]$ can be computed by directly calculating the overlaps in the definition in Eq.~\eqref{eq:C}, if the tangent space is a K\"{a}hler manifold (i.e. $J^2 = -1$, as discussed in Sec.~\ref{sec:NGSclosed}), then from the relation $J\indices{^\mu_\nu}\ket{v_\mu} = i \ket{v_\nu}$ we can relate $\Im[C_\mu]$ to the complex structure $J\indices{^\mu_\nu}$ and the gradient $\partial_{\mu} \langle H \rangle$ using
\begin{align}\begin{split}
2 \Im[C_\nu ] &= i\bra{v_\nu} H \ket{\psi} - i \bra{\psi} H \ket{v_\nu} \\ 
&= -\bra{v_\mu} (J\indices{^\mu_\nu})^T H \ket{\psi} - \bra{\psi} H J\indices{^\mu_\nu}\ket{v_\mu} \\ 
&= -\sum_\mu J\indices{^\mu_\nu} \partial_\mu E.
\end{split}\label{eq:ImC_J}
\end{align}
As such, the non-Hermitian equations of motion in Eq.~\eqref{eq:EoMs_Heff} can be computed from $g$, $\partial_\mu \langle H \rangle$, and $\partial_\mu \langle K \rangle$. The total number of elements to compute scales as $M^2 + 2M$, compared to $M^2 + M$ for purely real- or imaginary-time evolution, where $M$ is the number of variational parameters. 

\vspace{5mm} \noindent \textbf{Example: Computing $\pmb{ \Im[C_\mu] }$ for the coherent state ansatz.}
Here, we relate ${\rm Im}[C_\mu]$ to the energy gradients $\partial_\mu \langle H \rangle$ for the coherent state ansatz with explicit normalization and phase factors, $\ket{\psi} = e^{\kappa + i \theta} e^{-(x^2 + y^2)/2} e^{(x + iy)a^\dagger}\ket{0}$, with $\vec{z} = (\kappa,\theta,x,y) \in \mathbb{R}$. The tangent vectors corresponding to $\mu \in \{1,2,3,4\}$ are 
\begin{subequations}\begin{align}\label{eq:CohStateAnsatzTangentVectors}
    \ket{v_1} &= \ket{\psi}, \\ 
    \ket{v_2} &= i \ket{\psi}, \\ 
    \ket{v_3} &= (a^\dagger - x)\ket{\psi}, \\ 
    \ket{v_4} &= (i a^\dagger - y)\ket{\psi}. 
\end{align}\end{subequations}

We begin by relating $\Im[C_1] = \Im[\bra{\psi}H\ket{v_1}]$ to $\partial_{z_2} E$, 
\begin{align}\begin{split}
    \Im[C_1] &= -\frac{1}{2} [i\bra{\psi}H\ket{v_1} - i\bra{v_1}H\ket{\psi}] \\ 
    &= -\frac{1}{2} (\bra{\psi} H \ket{v_2} + \bra{v_2} H \ket{\psi}) = -\frac{1}{2} \partial_{z_2} E.
\end{split}
\end{align}
Similarly for $\Im[C_2] = \Im[\bra{\psi}H\ket{v_2}]$, we have
\begin{align}
    \Im[C_2] = \frac{1}{2} \partial_{z_1} E. 
\end{align}
To relate $\Im[C_3] = \Im[\bra{\psi}H\ket{v_3}]$ to the gradients of $E$, we notice that we can write $i \ket{v_3}$ as a \emph{real} span of the tangent vectors $\{v_\mu\}$. That is, $i \ket{v_3} = \ket{v_4} + y \ket{v_1} - x \ket{v_2} $. Using this we obtain
\begin{align}
      \Im[C_3] = -\frac{1}{2} \left( \partial_{z_4} E + y \partial_{z_1} E - x \partial_{z_2} E \right),
\end{align}
and finally for $\Im[\bra{v_4}H\ket{\psi}]$, using $i \ket{v_4} = - \ket{v_3} - x \ket{v_1} - y \ket{v_2}$, we obtain
\begin{align}
    \Im[\bra{v_4}H\ket{\psi}] = \frac{1}{2} \left( \partial_{z_3} E + x \partial_{z_1} E + y \partial_{z_2} E \right).
\end{align}
Therefore, we have related the computation of $\Im[C_\mu]$ to the energy gradients of $E = \langle H \rangle$. In this example, we explicitly noticed that $i\ket{v_\mu}$ can be written as a real span of tangent vectors $\{\ket{v_\mu}\}$. In the supplementary material~\cite{supp}, we provide the constructions of $g_{\mu \nu}$, $\omega_{\mu \nu}$ and $J\indices{^\mu_\nu}$ for this single-mode coherent state ansatz, showing that $\ket{v_\nu} = J\indices{^\mu_\nu} \ket{\mu}$, as derived in Eq.~\eqref{eq:ImC_J}. We provide $J\indices{^\mu_\nu}$ for the NGS ansatz and prove that $J^2 = -1$.

\begin{figure}
    \centering
    \includegraphics[width=\linewidth]{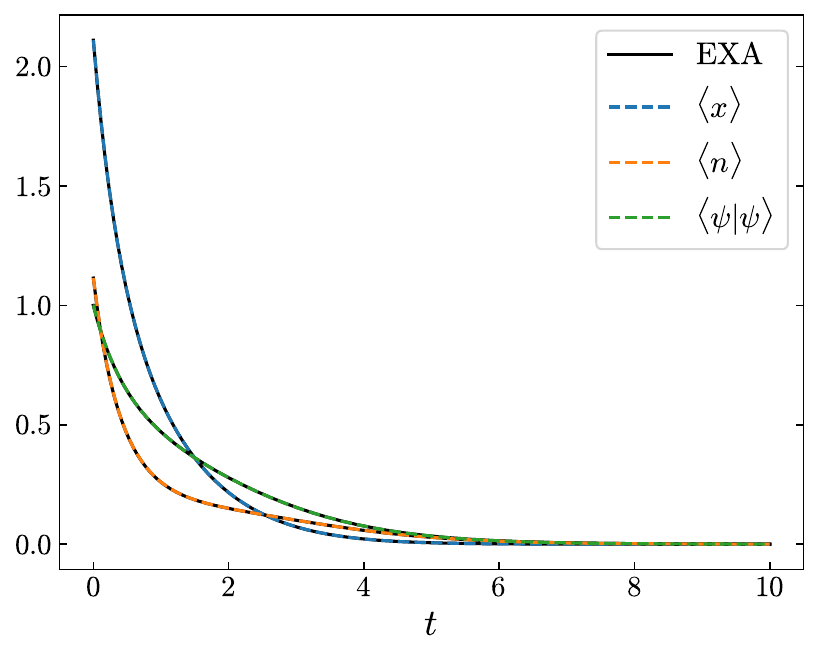}
    \caption{Dynamics of the effective Hamiltonian $H_{\rm eff} = \xi (a + a^\dagger) - (i/2) \kappa a^\dagger a$ with $\xi = 0.5$, $\kappa = 1.0$ using the NGS ansatz without squeezing (i.e. $\vec{\zeta} = 0$) with $N_p = 2$. The non-Hermitian term corresponds to a cavity loss jump operators. The initial state is a random superposition of two coherent states. The NGS results (dashed colours) agree perfectly with exact numerics (solid black), including the decay of the norm $\braket{\psi}$ due to the non-hermitian term.} 
    \label{fig:NonHermitianEvolution}
\end{figure}

Finally, we provide a simple example of non-Hermitian dynamics using NGS and the equations of motion in Eq.~(\ref{eq:EoMs_Heff}) by computing the evolution of the coherent state ansatz for $N_b = 1$ mode under $H_{\rm eff} = \xi (a + a^\dagger) - (i/2) \kappa a^\dagger a$. Since the Hamiltonian is a Gaussian operator, if the initial state of interest can be accurately described by the NGS ansatz with $N_p$ coherent states, NGS is exact in that it captures precisely the dynamics at all times also with $N_p$ coherent states. The results for an $N_p = 2$ initial state and a comparison against exact numerics are shown in Fig.~\ref{fig:NonHermitianEvolution}, depicting perfect agreement as expected.

\subsection{Quantum jumps}
\label{sec:QuantumJumps}
We will now incorporate the action of a quantum jump $c_m$ in the NGS formalism. After each quantum jump the wavefunction may (i) remain in the variational manifold, or (ii) leave it. We discuss these two possibilities using the examples of single particle loss and gain, respectively. We have chosen these two processes as they are often the dominant sources of single particle decoherence in a variety of systems. 
\\~
\subsubsection*{(i) Jumps inside the manifold} To demonstrate the effect of a jump operator that produces a state that remains within the variational manifold, we consider single particle loss at rate $\kappa^{(1)}$. The jump operator is $c = \sqrt{\kappa^{(1)}}a$. The action of $c$ on the single-mode NGS ansatz in Eq.~(\ref{eq:psi}) \emph{without} squeezing, i.e. $|\vec{\zeta}|=0 \; \forall \; p$, is given by, 
\begin{align}\begin{split}
    \sqrt{\kappa^{(1)}}a \ket{\psi} &= \sqrt{\kappa^{(1)}}a \sum_p e^{\kappa_p + i \theta_p} D(\alpha_p)\ket{0} \\ 
    &= \sum_p \sqrt{\kappa^{(1)}} e^{\kappa_p + i \theta_p} \alpha_p D(\alpha_p)\ket{0}  \\
    &= \sum_p e^{\kappa_p^\prime + i \theta_p^\prime} D(\alpha_p)\ket{0}, 
\end{split}\end{align}
where we used $a\ket{\alpha} = \alpha \ket{\alpha}$ and defined the updated norm and phase factors as
\begin{subequations}\begin{align}
    \kappa_p^\prime &= \log(\sqrt{\kappa^{(1)}} e^{\kappa_p} \abs{\alpha_p} ), \\ 
    \theta_p^\prime &= \arg(\abs{\kappa^{(1)}} e^{i \theta_p + \kappa_p} \alpha_p ).
\end{align}\end{subequations}
We can easily extend the above analysis to the two-photon loss case with the jump operator $c = \sqrt{\kappa^{(2)}} a^2$, where the updated norm and phase factors are now defined as 
\begin{subequations}\begin{align}
    \kappa_p^\prime &= \log(\sqrt{\kappa^{(2)}} e^{\kappa_p} \abs{\alpha_p}^2), \\ 
    \theta_p^\prime &= \arg(\sqrt{\kappa^{(2)}} e^{i \theta_p + \kappa_p} \alpha_p^2).
\end{align}\end{subequations}

It is important to note that our results rely on the coherent state being the eigenstate of the jump operator considered above. Thus, for any other state, e.g. the squeezed coherent state $\ket{\alpha, \zeta}$, the single-particle loss jump operator will take the state out of the variational manifold. This will be a more generic scenario for most jump operators. In the next section we describe how to deal with such situations.

\subsubsection*{(ii) Jumps outside the manifold}\label{sec:JumpsOutsideManifold}

If the state after the jump is not within the variational manifold, we project it back to the manifold by maximizing its fidelity with the variational ansatz. To do so we use gradient descent (GD) as an efficient numerical procedure. We note that while simulated annealing (SA) finds a global extremum of a function in the asymptotic limit of infinitely slow cooling rate, we find that in all cases studied in this work, the performance of GD is comparable to that of SA (the infidelity difference between the post-jump variational state found by each of the two methods is $\lesssim 10^{-3}$), with the advantage that GD is typically significantly faster.

\begin{figure}
    \centering
    \includegraphics[width=\linewidth]{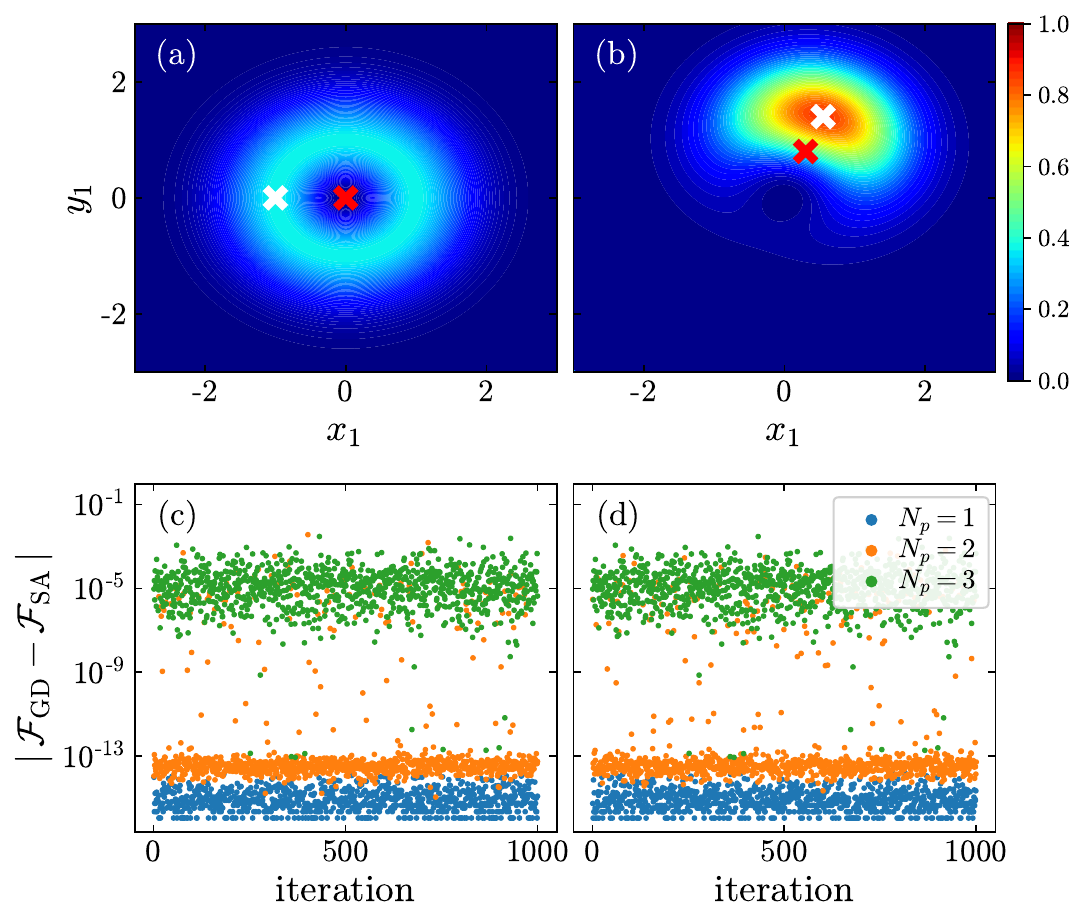}
    \caption{(a,b) Fidelity of $a^\dagger \ket{\alpha}$ projected back onto the variational manifold here formed by the set of all coherent states $|\tilde{\alpha} = \tilde{x} + i \tilde{y} \rangle$. The red cross denotes the initial state upon which acts the jump operator $a^\dag$, and the white cross denotes the maximum of Eq.~(\ref{eq:ab_overlap}) found via GD. The used initial states are (a) $\ket{\alpha=0}$ (with a ring of global maxima due to the symmetry of $a\ket{0}$) and (b) $\ket{\alpha = 0.3 + i 0.8}$. 
    (c,d) Difference in optimized fidelities, cf.~Eq.~(\ref{eq:fidelity}), found by GD vs.~bounded SA. The jump operators are (c) $a^\dagger$ and (d) $x=a+a^\dag$.} 
    \label{fig:fig_aop_Np1}
\end{figure}

Again, we consider a single mode NGS ansatz without squeezing (i.e. $\vec{\zeta} = 0$), which we write as 
\begin{align}
    \ket{\psi} = \sum_{i=1}^{N_p} c_i \ket{\alpha_i},
    \label{eq:psi_opt}
\end{align}
with complex coefficients $c_i$ and complex amplitudes $\alpha_i$. For single photon gain with jump operator $c=a^\dagger$, the state after applying the jump operator $a^\dagger \ket{\psi}$ is projected back onto a generic variational state $|\tilde{\psi}\rangle$ by  optimising its variational parameters $\tilde{c}_i,\tilde{\alpha}_i$ to maximize the normalised fidelity given by, 
\begin{align}
    \mathcal{F} = \frac{|\langle \tilde{\psi} | a^\dagger | \psi \rangle|^2}{{|\langle \tilde{\psi} | \tilde{\psi }\rangle|}{|\langle \psi | a a^\dagger | \psi \rangle|}},
    \label{eq:fidelity}
\end{align}
where the un-normalised overlap is 
\begin{align}
    \langle \tilde{\psi} | a^\dagger | \psi \rangle = \sum_{i,j}^{N_p} \tilde{c}_i^* c_j \tilde{\alpha}_i^* e^{- \frac{1}{2}(|\alpha|^2 + |\tilde{\alpha}|^2 ) + \tilde{\alpha}_i^* \alpha_j}, 
    \label{eq:ab_overlap}
\end{align}
and the normalisation factors are 
\begin{subequations}\begin{align}
    \bra{\psi}\ket{\psi} &= \sum_{i,j}^{N_p} c_i^* c_j e^{-\frac{1}{2} (|\alpha_i|^2 + |\alpha_j|^2 + \alpha_i^* \alpha_j)}, \\ 
\nonumber    \bra{\psi}aa^\dagger\ket{\psi} &= \sum_{i,j}^{N_p} c_i^* c_j (1 + \alpha_i^* \alpha_j) \\
    &\times e^{-\frac{1}{2} (|\alpha_i|^2 + |\alpha_j|^2) + \alpha_i^*  \alpha_j}. 
\end{align}\end{subequations}
In the case of $N_p = 1$, with $\alpha=|\alpha|{\rm e}^{i \varphi_\alpha}$, $\tilde{\alpha}=|\tilde{\alpha}|{\rm e}^{i \varphi_{\tilde{\alpha}}}$, Eq.~ \eqref{eq:ab_overlap} is maximized for $\varphi_{\tilde{\alpha}} = \varphi_\alpha$ and for amplitude of $\tilde{\alpha}$ which is the solution to $|\tilde{\alpha}|^2 - |\alpha|\,|\tilde{\alpha}|-1=0$. For instance if $\alpha=0$, $|\tilde{\alpha}|=1, \; \forall \varphi_{\tilde{\alpha}}$, cf. Fig.~\ref{fig:fig_aop_Np1}(a), and similarly for $\alpha \neq 0$, cf. Fig.~\ref{fig:fig_aop_Np1}(b). The starting point of the GD search is denoted by the red cross and the maximum of the numerically found maximum of the overlap, Eq.~(\ref{eq:ab_overlap}), by the white cross. We remark that the achievable fidelity after projecting back to the manifold is strongly dependent on the jump operator and the number of coherent states. For instance, for the case shown in Fig.~\ref{fig:fig_aop_Np1}(a), the state after the jump is a Fock state $\ket{1}$. As such, after projecting it back to a single coherent state, its fidelity is given by  ${\rm e}^{-|\tilde{\alpha}|^2/2} |\alpha|^n/\sqrt{n!} \rightarrow {\rm e}^{-1/2} \approx 0.61$ with $n=1$ and $|\tilde{\alpha}|=1$.

In the case of the NGS ansatz with $N_p > 1$, the optimization landscape becomes more complex. In Fig.~\ref{fig:fig_aop_Np1}(c)-(d), for $N_p =1$ (blue), $N_p = 2$ (orange) and $N_p=3$ (green) we use Eq.~\eqref{eq:fidelity} and plot the difference in optimized fidelities $|{\cal F}_{\rm GD} - {\cal F}_{\rm SA}|$, with the optimization performed by GD with backtracking (${\cal F}_{\rm GD}$) and bounded SA (${\cal F}_{\rm SA}$). 
We consider two quantum jumps, (c) single-particle gain $c = a^\dagger$ and (d) momentum kicks $c = x = a + a^\dag$. Our choice of single-particle gain is motivated by its relevance to many spin-boson systems (see Secs.~\ref{sec:Results} and \ref{sec:conclusions}), whilst the momentum kick jump plays a crucial role in laser cooling large ion crystals~\cite{Gerritseninprep}. We generate $10^3$ initial (pre-jump) random states $\ket{\psi}$. As the generic variational starting state $|\tilde{\psi}\rangle$ to be optimized, for GD we use the pre-jump state $\ket{\psi}$, whilst for the SA we seed a random starting state $|\tilde{\psi}\rangle$ whose coefficients (see Eq.~(\ref{eq:psi_opt})) are drawn from a uniform unit distribution $\{c_i,\alpha_i,\tilde{c}_i,\tilde{\alpha}_i\} \in [0,1]$. We set a sufficiently slow SA cooling rate such that the algorithm converges to the same local maxima irrespective of the randomly chosen starting point. For all the studied cases, the fidelities of the states obtained by the two numerical optimizers agree within $\lesssim 10^{-3}$. 

\subsection{Time evolution}
\label{sec:Trottetisation}

We are now equipped to implement the quantum trajectories program for the NGS ansatz outlined in Sec.~\ref{sec:QuantumTrajectories}. In principle, one could evolve the wavefunction $\ket{\psi}$ between the jumps according to the equations of motion in Eq.~(\ref{eq:EoMs_Heff}) while tracking the decay of the norm to identify the time $t_j$ of a jump. In practice we find it convenient to Trotterize the time evolution between jumps governed by $H_{\rm eff}$, Eq.~(\ref{eq:Heff2}), in the usual way as
\begin{align}
    e^{-i(H-iK)\delta t} \approx e^{-iH\delta t} e^{-K \delta t} 
\end{align}
for sufficiently small $\delta t$. The norm during the unitary dynamics under ${\rm e}^{-iH\delta t}$ is preserved due to the use of the norm factors $\kappa$ as variational parameters (we note that the opposite case, namely the absence of a global norm factor as a variational parameter, can result in unphysical couplings, see Ref.~\cite{Hackl_2020_SciPost}). During the imaginary-time evolution ${\rm e}^{-K \delta t}$ we track the decay of the norm to determine the time of the jump $t_j$, at which point we apply the corresponding jump operator $c_m$. 

\section{Truncated Wigner Approximation for spins and bosons}\label{sec:twa}

The phase space picture of quantum mechanics provides alternative means to simulate and analyze the quantum many-body dynamics in systems with mixed spin and bosonic degrees of freedom, in particular in a semi-classical framework known as the truncated Wigner approximation (TWA)~\cite{Polkovnikov2010}. The Wigner-Weyl transform maps Hilbert space operators $O$ of a quantum system to functions of classical phase space variables, known as Weyl symbols $O_W$. The Weyl symbol corresponding to the density matrix is known as the Wigner function and provides a full ensemble description of arbitrary quantum states in terms of a (potentially negative) quasi-probability distribution.

A general Wigner-Weyl transform can be defined using the framework of phase point operators~\cite{wootters1987a}. For example, considering particles in 1D with positions $\mathbf{x}$ and momenta $\mathbf{p}$, operators $\mathcal{A}(\mathbf{x},\mathbf{p})$ for each point in phase space define the Wigner-Weyl transformation via $O_W(\mathbf{x},\mathbf{p}) = {\rm tr}[\mathcal{A}(\mathbf{x},\mathbf{p}) O]/\sqrt{2\pi}$ and vice versa $O = \int d\mathbf{x} d\mathbf{p}\, O_W(\mathbf{x},\mathbf{p}) \mathcal{A}(
\mathbf{x},\mathbf{p})$. Given a proper orthonormal definition of phase point operators~\cite{wootters1987a}, for any state $\rho$ and any observable $Q$, expectation values can be evaluated from the Wigner function $W(\mathbf{x},\mathbf{p}) = {\rm tr}[\mathcal{A}(\mathbf{x},\mathbf{p}) \rho]/\sqrt{2\pi}$ via $\langle Q \rangle = \int d\mathbf{x} d\mathbf{p}\, Q_W(\mathbf{x},\mathbf{p}) W(\mathbf{x},\mathbf{p})$. Equivalent constructions can be made for spin phase spaces, either using spin-boson mappings (suitable for large spin scenarios)~\cite{Polkovnikov2010}, using spherical coordinate representations of spins $A(\theta, \phi)$~\cite{FleischhauerDTWA}, or for phase spaces using only a discrete set of points~\cite{wootters1987a}.

Closed-system time-evolution equations of motion can be obtained by Wigner-Weyl transforming the Heisenberg equations of motion, which leads to the exact quantum dynamics for Weyl symbols being governed by $\dot Q_W(\mathbf{x},\mathbf{p}) = \{Q_W(\mathbf{x},\mathbf{p}), H_W(\mathbf{x},\mathbf{p})\}_{\rm MB}$, using the Weyl symbol of the Hamiltonian $H_W$, and the Moyal bracket defined as $\{Q_W, H_W\} = 2 Q_W \sin(\mathbf{\Lambda/2}) H_W$, with $\mathbf{\Lambda}$ the symplectic operator (with $\hbar\equiv 1$) $\mathbf{\Lambda}  = \sum_{i} \frac{\overleftarrow{\partial}}{\partial x_i} \frac{\overrightarrow{\partial}}{\partial p_i} - \frac{\overleftarrow{\partial}}{\partial p_i} \frac{\overrightarrow{\partial}}{\partial x_i}$. Expanding the sine function in the Moyal bracket at the lowest order is know as TWA and leads to a classical evolution of Weyl symbols $\dot Q_W(\mathbf{x},\mathbf{p}) \approx \{Q_W(\mathbf{x},\mathbf{p}), H_W(\mathbf{x},\mathbf{p})\}_{\rm P}$, where $\{\cdot,\cdot\}_{\rm P}$ now denotes the classical Poisson bracket. The Poisson bracket ensures that the Weyl symbols for any complex observable will always factorize into phase space variables, and therefore in TWA it suffices to only compute the classical evolution of the phase space variables~\cite{zhu2019a}. This makes TWA a very practical and efficient numerical method for the case of a positive initial Wigner function: Random positions and momenta can be sampled from the Wigner function and evolved in parallel using classical equations of motion giving $\mathbf{x}_{\eta}(t)$ and $\mathbf{p}_{\eta}(t)$ for trajectory $\eta$. Expectation values in TWA are then statistically approximated by $\langle Q \rangle \approx \frac{1}{n_{\rm traj}}\sum_\eta^{n_{\rm traj}} Q_W(\mathbf{x}_{\eta}(t), \mathbf{p}_{\eta}(t))$, using $n_{\rm traj}$ trajectories.

Importantly, for small-spin systems, and in particular for spin-1/2 models as considered here, TWA can be drastically improved when using a sampling of the initial Wigner function using only a discrete set of initial phase points \cite{Schachenmayer2015a}. Considering a system consisting of a single spin-1/2 described by the Pauli operators $\bm{\sigma}=\left(\sigma_x, \sigma_y, \sigma_z\right)$, we define the corresponding phase space variables as ${\mathcal{S}}= \left(S_x, S_y, S_z\right)$. One can then define discrete Wigner functions which are only defined for for the 8 different discrete points~$\mathbf{\mathcal{S}}_0 = (\pm 1, \pm 1, \pm 1)$. For example, taking a state of $N_s$ spin-1/2 particles of the form $\ket{\psi} = \bigotimes_{i=1}^{N_s} \ket{\downarrow}_i$, it is straightforward to show that any possible observable can be exactly described by sampling each spin from a discrete Wigner function with the only non-zero values of $W^\downarrow_i(S_x = \pm 1, S_y = \pm 1, S_z = -1) = 1/4$. Correspondingly, the state $\ket{\uparrow}_i$ is exactly described by the discrete distribution with non-zero elements $W_i^\uparrow(S_x = \pm 1, S_y = \pm 1, S_z = +1) = 1/4$. 

Furthermore, it can be shown in general that equivalent discrete sampling strategies can lead to exact quantum state descriptions 
for general discrete $D$-level systems and for eigenstates of general spin-$S$ operators, in the sense that the measurement statistics for any observable can be exactly reproduced from sampling the Wigner function~\cite{zhu2019a}. Discrete sampling in combination with classical evolution is known as (generalized) discrete truncated Wigner approximation, (G)DTWA~\cite{Schachenmayer2015a, zhu2019a}. Classical equations of motion for the spin-variables can be derived by Wigner-Weyl transforming the Heisenberg equations of motion while factorizing the Weyl symbols into the phase space variables. (G)DTWA has been shown to capture quantum features in spin-model dynamics in several theory settings~\cite{Schachenmayer2015b, acevedo2017exploring, kunimi2021performance, Perlin2020,Muleady2023}, and in comparison with experiments~\cite{lepoutre2019out, orioli2018relaxation,Franke2023,Alaoui2024}. 

Below we will consider a system consisting not only of spins but also of bosonic $ a \left(a^\dagger\right)$ degrees of freedom. For the bosonic part we will consider the complex numbers $ a\rightarrow A$ and $a^\dagger \rightarrow A^*$ as the classical phase space. We note that for additional bosonic degrees of freedom with operators denoted as $b \left(b^\dagger\right)$ one can introduce a corresponding classical phase space with $b \left(b^\dagger\right) \to B\left(B^*\right)$ (see for example Sec.~\ref{sec:Results}). Then, considering a system with  $N_s$ spin-1/2 particles coupled to $N_{a/b}$ bosonic modes $a/b$, computing expectation values of an observable $Q(\bm{\sigma}_i, a_j, b_k)$ with TWA at time $t$ corresponds to numerically evaluating 
\begin{align}
    \langle O&\left(t; \{\bm{\sigma}_i,  a_j, b_k\}\right)\rangle \approx \int \prod_{i=1}^{N_s}  \prod_{j=1}^{N_a} \prod_{k=1}^{N_b} d\mathbf{\mathcal{S}}_0^i d^2 A_0^j d^2 B_0^k \\ 
  & W_i(\mathbf{\mathcal{S}}_0^i) W_j(A_0^j) W_k(B_0^k) 
  O_W\left(\{\mathbf{\mathcal{S}}_{\rm cl}^i(t), {A}_{\rm cl}^j(t), {B}^k_{\rm cl}(t)\}\right), \nonumber
\end{align}
where $d\mathbf{\mathcal{S}}_0=dS_0^xdS_0^ydS_0^z$, $d^2 A_0=d {\rm{Re}} A_0 d{\rm{Im}} A_0 /\pi$, and the subscript $0$ indicates the initial values at $t=0$. The classical variables for spin $i$ and bosons $j$ and $k$ are sampled from the initial Wigner functions $W_i(\mathbf{\mathcal{S}}_0^i)$,  $W_j(A_0^j)$, and $W_k(B_0^k)$, respectively. Note that we always assume an initial product state between all degrees of freedoms such that the Wigner functions factorize.  For the spins we will use the discrete distributions $W^{\uparrow/\downarrow}_i$ defined above, while for the bosonic modes we use standard continuous Wigner functions, in particular
\begin{equation}
    W_j(A_0^j)=\frac{1}{2\pi w_j^2}\exp{-\abs{A_j-\bar{\alpha}_j}^2/(2 w_j^2)},
    \label{eqn:DTWA_kernel}
\end{equation}
where $w_j^2=(\bar{n}_j+1/2)/2$ and $\alpha$ center of the Wigner function. For the vacuum state $\bar{n}_j=\bar{\alpha}_j=0$, while for a coherent state $\ket{\alpha}$, $\bar{n}_j=0$ and $\bar{\alpha}_j=\alpha$ (see Refs.~\cite{Polkovnikov_2010_AnnPhys, Orioli2017} for more details). $O_W\left(\{\mathbf{\mathcal{S}}_{\rm cl}^i(t), {A}_{\rm cl}^j(t), {B}^k_{\rm cl}(t)\}\right)$ is the Weyl symbol corresponding to the observable of interest. We use the subscript \emph{cl} on the time-dependent variables $\mathbf S^i_{\rm cl}(t)$, $A^j_{\rm cl}(t)$, and $B^k_{\rm cl}(t)$ to indicate that they obey the classical equations of motion for spin $i$, boson $j$, and boson $k$, respectively. In Sec.~\ref{sec:Results} below we will provide the full classical equations of motion for our problem of interest for examples of both closed and open system dynamics. 
(G)DTWA methods have been recently developed further to also include open-system dynamics under Lindblad master equations~\cite{Huber2021,Huber2022,Singh2022,FleischhauerDTWA,Mink2023}. For our simulations we follow the procedure in  Ref.~\cite{FleischhauerDTWA} and use a spherical coordinate parametrization of the phase space for spin $i$ with phase point operators defined as
\begin{equation}
    \label{eqn:spinWignerkernel}
    \mathcal{A}_i(\theta_i,\phi_i)= \frac{1}{2}[\mathbb{1}_i+ \mathbf{s}(\theta_i,\phi_i) \cdot \bm{\sigma}_i],
\end{equation}
where we use the vector on the surface of a sphere with radius $\sqrt{3}$, 
$\mathbf{s}_i=\sqrt{3} \left(\sin{\theta_i} \cos{\phi_i}, -\sin{\theta_i} \sin{\phi_i}, -\cos{\theta_i}\right)^\intercal$. 

In~\cite{FleischhauerDTWA} it was shown that for open spin-1/2 models, it is convenient to work with flattened Wigner functions of the form $\chi_i(\theta_i, \phi_i) \equiv W_i(\theta_i, \phi_i) \frac{\sin\theta_i}{2\pi}$. The equations of motion for an open system can then be found by deriving correspondence rules (reminiscent of Bopp representations for for bosonic systems~\cite{Polkovnikov2010}), i.e.~rules for mapping terms such as $X_i \rho_i$, $\rho_i X_i$, or $X_i \rho_i X_i^\dag$ with Pauli operators $X = \sigma_i^{x,y,z}$ to the phase space, which leads to terms incorporating the four linearly independent differential operations $\mathbb{1}, \frac{d}{d\theta_i},\frac{d}{d\phi_i}, \frac{d^2}{d \phi_i^2}$ acting on $\chi(\theta_i, \phi_i)$. It can be shown that the resulting EOMs lead to standard Fokker-Planck equations. This is only valid, without further approximations, for systems of non-interacting spins or if the initial state is a large coherent spin state. In these scenarios, the dynamics are given by the solution to the Fokker-Planck equations. Rather than solving these equations directly, we solve the corresponding Itô stochastic differential equations and average the relevant expectation values over many trajectories \cite{Gardiner_2004_Book}.

In our discrete sampling we select the initial angles $\theta_i,\phi_i$ according to the parametrization given in Eq.~\eqref{eqn:DTWA_kernel}. 
However, rather than sampling from the discrete $W_i^{\uparrow / \downarrow}$ Wigner distribution, we sample from a slightly modified flattened Wigner function 
\begin{equation}
    \chi^{ \uparrow /\downarrow}(\theta,\phi)    = \frac{1}{2\pi}\delta(\theta \pm \arccos{\frac{1}{\sqrt{3}}}),
\end{equation}
which is generated by rotating the discrete Wigner function $W_i^{\uparrow / \downarrow}$ around the z-axis. This initial Wigner function is uniform in $\phi$ and thus guarantees that $\frac{\partial^2}{\partial \phi_m \partial(\theta_n-\theta_{n+1})}$ cross diffusion terms, which are dropped in TWA, vanish at $t=0$ \cite{FleischhauerDTWA}. For more details, we refer the reader to Ref.~\cite{FleischhauerDTWA} and Sec.~\ref{sec:Results} where the detailed equations for our problem of interest are introduced.

\section{Results}
\label{sec:Results}

\begin{figure}
    \centering
    \includegraphics[width=0.75\linewidth]{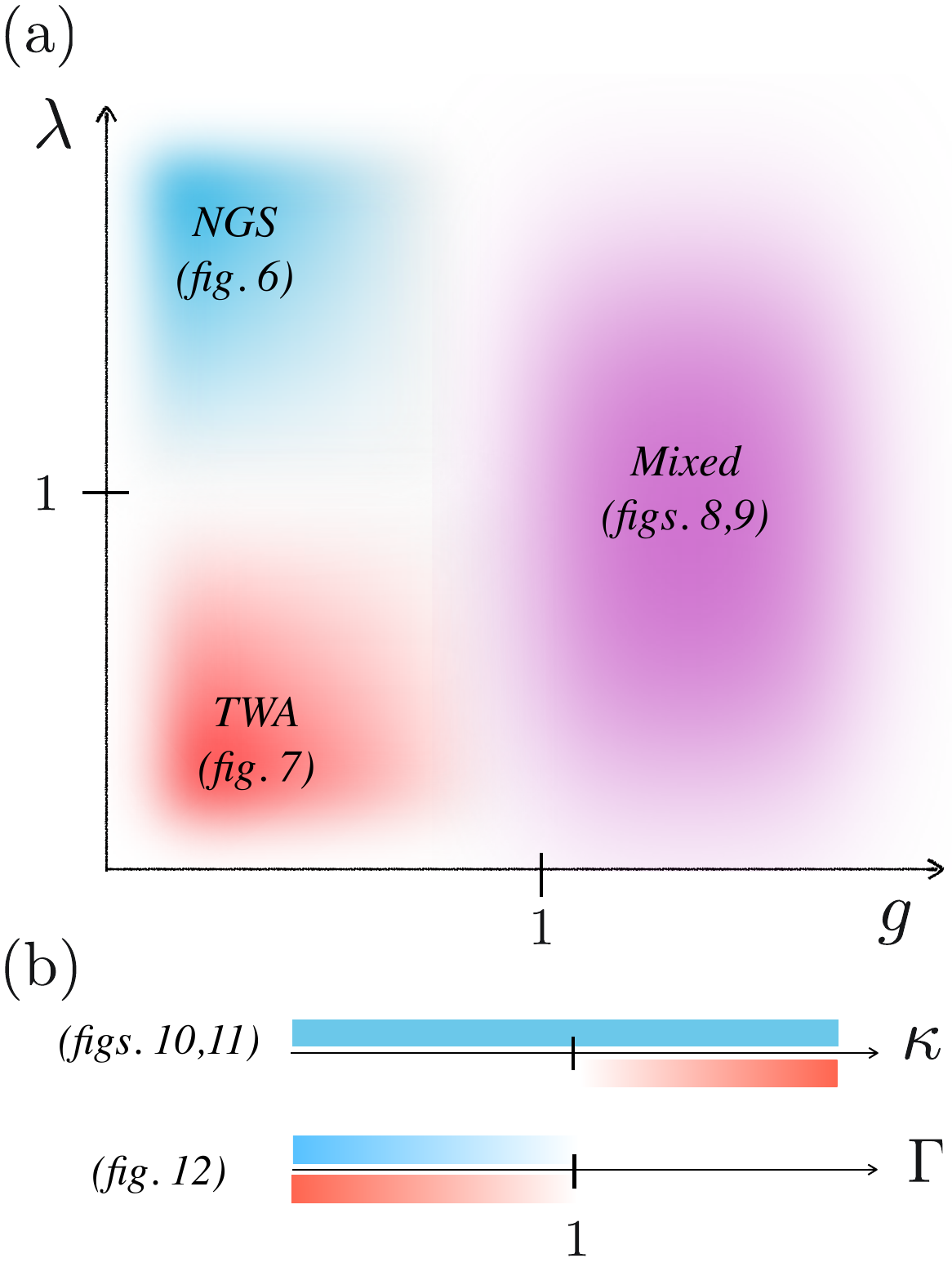}
    \caption{Schematic of the performance of TWA and NGS for (a) closed and (b) open dynamics. In the present study NGS is limited to at most $N_p = 16$ multi-mode coherent states and no squeezing.    
    (a) The performance in $g-\lambda$ parameter space for closed dynamics.
    When $g,\lambda \lesssim 1$, TWA performs well, whilst when $g \lesssim 1, \lambda \gtrsim 1$, NGS is the better choice. When $g \gtrsim 1$, TWA captures short-time dynamics, but can produce incorrect mid- to late-time results. In comparison, NGS typically does not capture quantitative details beyond the first spin relaxation, but does provide qualitative insights into the dynamics by correctly capturing the magnitude of the persistent spin-cavity dynamics. (b) Open dynamics. We operate in the $g \lesssim 1$, $\lambda \gtrsim 1$ region, so NGS outperforms TWA at small $\kappa$. At large $\kappa$, TWA performs well again, with both NGS and TWA agreeing. 
    When evolving under only $H_{\rm TC}$ with the spins decaying collectively at rate $\Gamma$, NGS and TWA agree when using the large $N_s$ Holstein-Primakoff (HP) transformation at small $\Gamma$. At large $\Gamma$ the collective spin quickly decays to the ground state, which is inaccessible as the first order expansion of the root in HP is no longer sufficient near the ground state since it was expanded around the excited state (small $a_s^\dagger a_s$), so neither method is able to capture intermediate- to late-time dynamics.}
    \label{fig:schem}
\end{figure}

To evaluate the performance of our two numerical methods, we consider the disordered Holstein-Tavis-Cummings model describing a system of $N_s$ spins coupled to a single bosonic mode ($a$), representing a coupling to a cavity mode, and $N_s$ local vibrational modes ($b_i$) associated with each spin. The system is described by the Hamiltonian \cite{Wellnitz_2022CommPhys}:
\beq
    H = H_{\rm TC} + H_{\rm vib} + H_{\rm H} + H_{\rm dis},
    \label{eq:H_molecule}
\eeq
where
\begin{subequations}
    \begin{align}
        H_{\rm TC} &= \frac{\Delta}{2} \sum_{j=1}^{N_s} (\sigma^z_j + \mathbb{1}) + \frac{g}{\sqrt{N_s}} \sum_{j=1}^{N_s} a \sigma_j^+ + a^\dag \sigma_j^-, \\
        H_{\rm vib} &= \nu \sum_{j=1}^{N_s} b^\dag_j b_j, \\
        H_{\rm H} &= -\frac{\lambda\nu}{2} \sum_{j=1}^{N_s} (b_j + b_j^\dag) (\sigma_j^z + \mathbb{1}), \\
        H_{\rm dis} & = - \frac{1}{2} \sum_{j=1}^{N_s} \epsilon_j (\sigma^z_j + \mathbb{1}),
    \end{align}
\end{subequations}
where $\Delta$ describes the detuning of the spin transition frequency relative to the cavity mode, $g/\sqrt{N_s}$ the single-spin coupling to the cavity, $\nu$ the frequency of the vibrational modes, $\lambda$ the relative strength of the Holstein coupling, and $\epsilon_j$ the disorder in the transition frequency for spin $j$. 

The dynamics of the Holstein-Tavis-Cummings model, in particular in the presence of disorder, has importance e.g.~in the field of polaritonic chemistry~\cite{herrera2016cavity}. It has been previously studied using a matrix product state method~\cite{Wellnitz_2022CommPhys}, and also using a similar non-Gaussian state framework to the one discussed here~\cite{Sun_2022_JChemPhys}. By tuning the relative strength of the various terms, this Hamiltonian can be reduced to spin-boson Hamiltonians applicable e.g.~to trapped ion quantum simulators, impurity models, and quantum chemistry. The relative strength of $g,\lambda,$ and $\nu$ allows us to go from the weak coupling regime between the spin and bosonic degrees of freedom, to a model governed by a Tavis-Cummings type interaction, to a Holstein coupling, or a combination thereof. We schematically illustrate the performance of the two numerical methods in each of these regimes in Fig.~\ref{fig:schem}(a). 

Furthermore, we investigate how both methods perform in the presence of sources of decoherence. In this case, we are interested in the evolution of the density matrix $\rho$ as described by the Lindblad master equation given in Eq.~\eqref{eq:master}. We study open dynamics with the following types of Lindblad jump operators: 
\begin{itemize}
    \item Cavity decay (rate $\kappa)$: $c_m=\sqrt{\kappa} a$,
    \item Single spin decay (rate $\gamma$): $c_m = \sqrt{\gamma}\sigma^-_i$, and 
    \item Collective spin decay (rate $\Gamma$): $c_m = \sqrt{\Gamma}\sum_i \sigma^-_i$.
\end{itemize}
We summarize the performance of the two methods in the presence of these decoherence sources schematically in Fig.~\ref{fig:schem}(b).

\subsection{Details of NGS and TWA simulations}
In principle the NGS ansatz can be used directly to treat the spin degrees of freedom. However, to avoid the explicit exponential scaling with $N_s$, we use a Holstein-Primakoff (HP) transformation to map each spin-$1/2$ to a bosonic mode. We use the following form of the HP transformation \cite{Vogl_2020_PRR},
\begin{subequations}
\label{eq:HP_Resum}
    \begin{align}
    S_+ &= (1-a_s^\dagger a_s) a_s, \\ 
    S_z &= 1/2 - a_s^\dagger a_s, 
\end{align}
\end{subequations}
which is exact for spin-$1/2$, and where we use the $s$ subscript to denote a bosonic operator acting on the spin degree of freedom. As such, the dynamics are restricted to the $\ket{0} \equiv \ket{e}$ and $\ket{1} \equiv \ket{g}$ subspace, with $\ketbra{g} = a_s^\dagger a_s$ and $\ketbra{e} = 1 - a_s^\dagger a_s$. The vacuum is a Gaussian state and can therefore be described using only $N_p = 1$. The Fock state $\ket{1}$ can be described using $N_p = 2$, by $\ket{1} = 1/\mathcal{N} \lim_{\alpha \rightarrow 0}(\ket{\alpha} - \ket{-\alpha})$, where $\mathcal{N}$ is a normalization factor~\cite{Marshall:23}, with $\alpha \sim 0.001$ sufficient to describe the state with high fidelity. As such, each spin can be captured using $N_p = 3$ coherent states. 

For the NGS simulations, we employ the NGS ansatz with $N_b = 2N_s + 1$ modes. We include cavity decay at strength $\gamma$ using the simple parameter-update prescription outlined in Sec.~\ref{sec:Results}. Note that as a consequence of the HP mapping, the ansatz is not well-suited to some scenarios. For example, there is no spin-spin coupling when evolving under only $H_{\rm dis}$, so an unentangled initial state will evolve as a tensor product of single spins, each requiring $N_p = 3$. The number of coherent states therefore scales exponentially in the number of spins, $3^{N_s}$. This limitation could potentially be mitigated by modifying the ansatz to be a superposition of squeezed displaced Fock states~\cite{Schlegel_2023}. 

For the TWA simulations, to more accurately capture the dynamics of the cavity and vibrational modes, we extend the set of TWA equations by including the classical equations of motion for the mode excitation numbers $a^\dagger a \rightarrow N_A,b_k^\dagger b_k \rightarrow N_{B_k}$. 
Including all three sources of decoherence described above, the stochastic equations of motion for the spin degrees of freedom are,
\begin{subequations}
\label{eqn:TWA_EOM}
\begin{align}
    \begin{split}
        \dot{\theta}_{i,\rm cl} &= (\Gamma+\gamma) ( \cot \theta_{i,\rm cl} - \frac{\csc \theta_{i,\rm cl} }{\sqrt{3}} ) \\
        &+ \frac{2g}{\sqrt{N_s}} \Im[e^{-i \phi_{i,\rm cl}} A_{\rm cl}]\\
        &- \frac{\Gamma\sqrt{3}}{2} \sum_j  \cos{(\phi_{i,\rm cl}-\phi_{j,\rm cl})}\sin{\theta_{j,\rm cl}}, 
     \end{split} \\ 
     \begin{split}
        d \phi_{i,\rm cl} &= \big( -\Delta +\epsilon_n  + 2 \lambda \nu \Re[B_{i,\rm cl}]\\
        & - \frac{2g}{\sqrt{N_s}}  \cot{\theta_{i,\rm cl}} \Re[A_{\rm cl} e^{-i \phi_{i,\rm cl}} ]\\
        &-\frac{\Gamma \sqrt{3}}{2} \sum_j \cot{\theta_{i,\rm cl}} \sin{\theta_{j,\rm cl}} \sin{(\phi_{i,\rm cl}-\phi_{j,\rm cl})} \big) dt \\
        & +  \sqrt{(\Gamma+\gamma) f(\theta_{i,\rm cl})} dW_{\phi_{i}}, 
     \end{split} 
\end{align}
\end{subequations}
 where we introduced the function $f(\theta_{i,\rm cl})=1 + 2 \cot{\theta_{i,\rm cl}}^2 - 2 \cot{\theta_{i,\rm cl}} \csc{\theta_{i,\rm cl}}/\sqrt{3}$. 
The stochastic behaviour of the equations of motion is generated by the Wiener increments $dW_{\phi_{i}}$ acting on $\phi_i$. For each timestep $dt$,
the Wiener increment $dW_{\phi_i}$ for each angle is independently drawn from a normal distribution with zero mean and a variance of $dt$.

The equations of motion for the bosonic degrees of freedom are given by
\begin{subequations}
\begin{align}
    \dot{A_{\rm cl}} &= -\frac{\kappa}{2}A_{\rm cl} -i \frac{\sqrt{3}g}{2\sqrt{N_s}} \sum_i e^{i \phi_{i,\rm cl}} \sin{\theta_{i,\rm cl}}, \\
    \dot{N}_{A,\rm cl} &= -\kappa N_{A,\rm cl} - \frac{\sqrt{3}g}{\sqrt{N_s}} \sum_i\sin{\theta_{i,\rm cl}} \Im[A_{\rm cl} e^{-i\phi_{i,\rm cl}}], \\
    \dot{ B}_{k,\rm cl} &= -i \nu B_{k,\rm cl} + i\frac{ \lambda \nu}{2} \left(1-\sqrt{3} \cos{\theta_{k,\rm cl}}\right), \\
    \dot{ N}_{B_{k,\rm cl}} &= \lambda \nu \left( 1- \sqrt{3} \cos{\theta_{k,\rm cl}}\right) \Im[B_{k,\rm cl}].
\end{align}
\end{subequations}

The initial state used for all NGS simulations includes a small amount of randomness for each variational parameter to break the degeneracy of the Gaussian states. We draw random values from a uniform distribution between $(0,10^{-4})$. We find that this is sufficient to ensure each Gaussian state in the superposition evolves independently, whilst preserving extremely small infidelity with the true initial state, $1-\mathcal{F} (t=0)< \mathcal{O}(10^{-6})$, as seen in Fig.~\ref{fig:Ns1_FidelityPlot} of the supplementary material~\cite{supp}. 

\subsection{Closed system dynamics}

In this section we compare the performance of the two numerical methods. We consider a system with $N_s=3$ spins, the corresponding three phonon modes, and one cavity mode. For this system size and suitable initial states and Hamiltonian parameters, choosing a moderate Fock state truncation $\sim 10$ allows us to compare our results to exact numerics. Our initial state is spins polarized up, the vibrational modes in the vacuum, and the cavity in a coherent state, $|\psi(0)\rangle=\vert {\uparrow}\rangle ^{\otimes N_s} \vert \alpha=1 \rangle_a \vert 0 \rangle_b^{\otimes N_s}$. Note in contrast to Ref.~\cite{Sun_2022_JChemPhys} our initial state is a superposition of several excitation manifolds precluding further simplifications to the ansatz. 

\begin{figure*}
    \centering
    \includegraphics[width=0.9\linewidth]{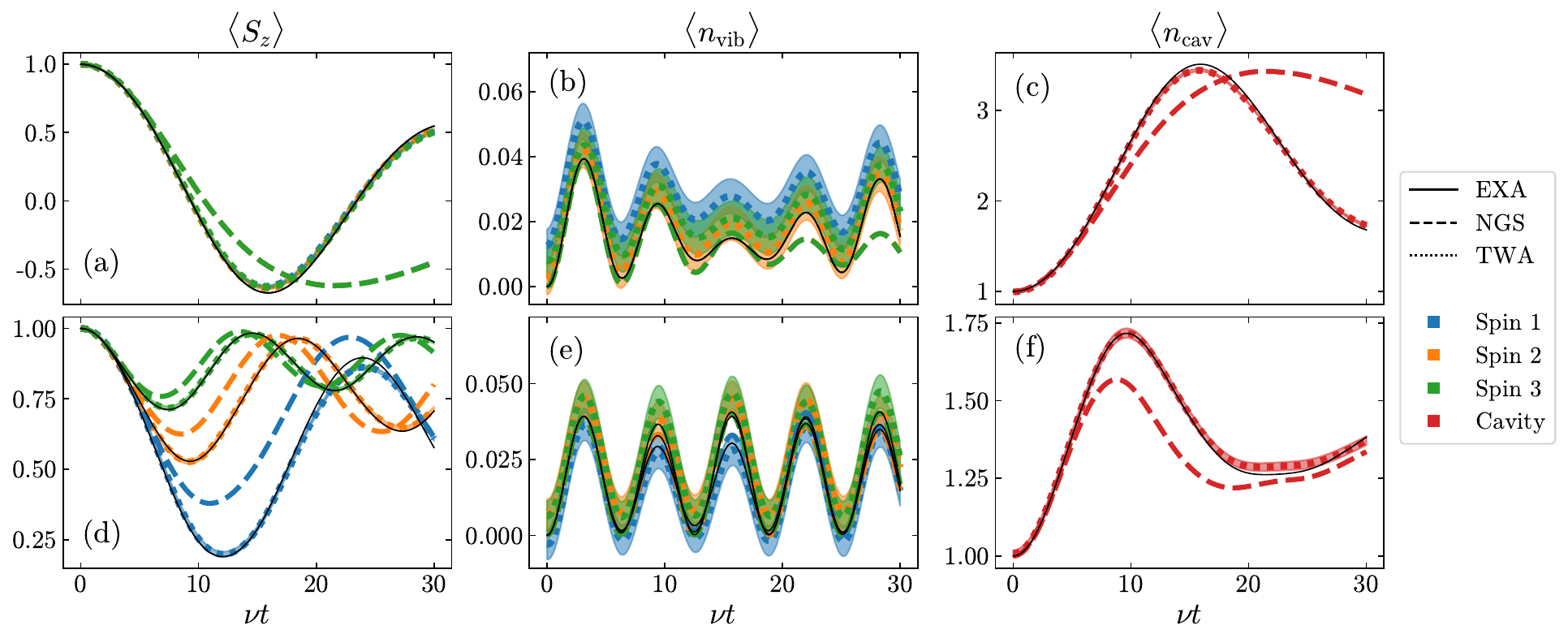}
    \caption{\emph{Closed dynamics}: $g = \lambda = 0.1$, $\Delta = 0$. \emph{Top row}: without disorder. Both NGS and TWA capture the initial spin relaxation, but NGS incorrectly predicts a slower revival. \emph{Bottom row}: disordered, $\vec{\epsilon} = [2g,3g,4g]$. Here, TWA captures the dynamics more accurately compared to NGS. NGS is with $N_p = 4$, TWA is with $n_{\rm traj} = 10^4$ with standard error shaded.}
    \label{fig:Data_Ns3_Region_g=0.1_λ=0.1.pdf}
\end{figure*}
\begin{figure*}
    \centering
    \includegraphics[width=0.9\linewidth]{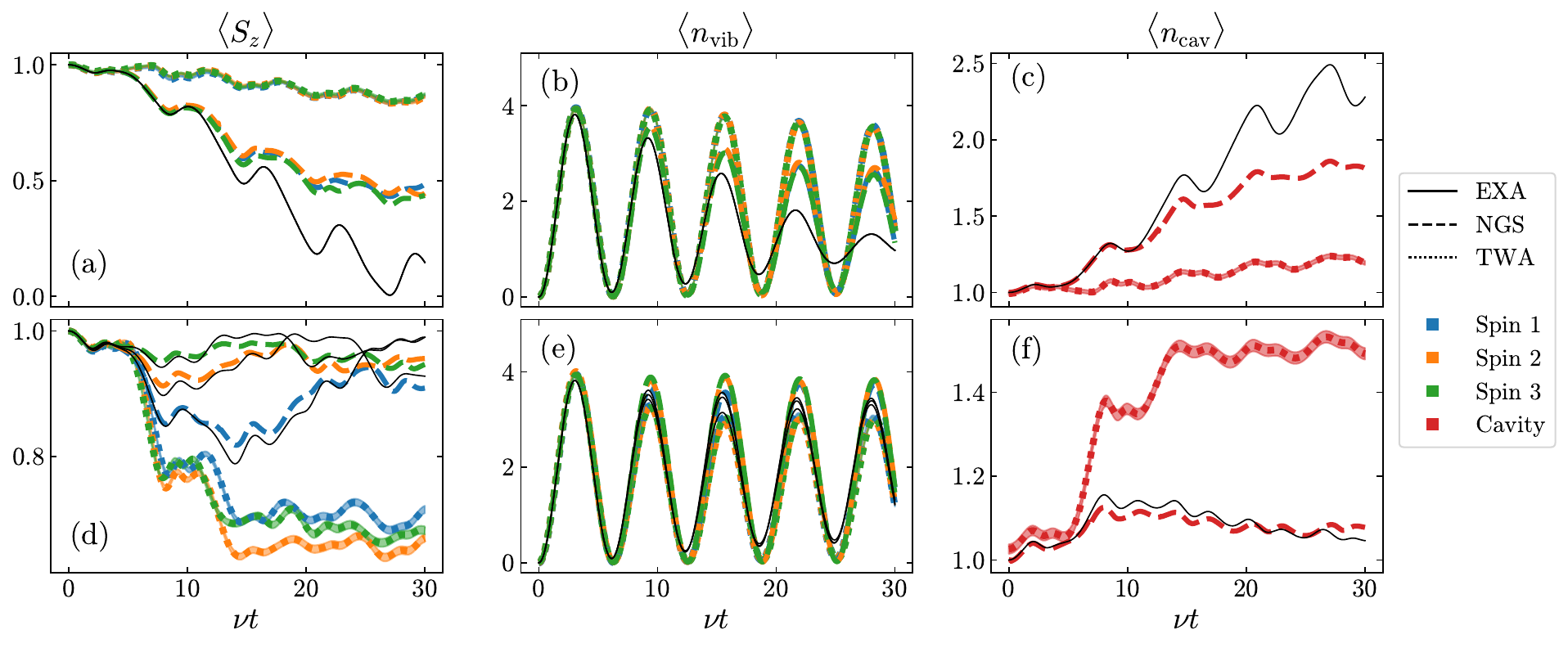}
    \caption{\emph{Closed dynamics}: $g = 0.1, \lambda = 1$, $\Delta = 0$. \emph{Top row}: without disorder. Neither TWA nor NGS completely capture the spin-cavity observables, but NGS more closely tracks the dynamics \emph{Bottom row}: disordered, $\vec{\epsilon} = [2g,3g,4g]$. NGS performs well, capturing all dynamics with small error. Interestingly, here TWA over estimates the magnitude of changes in the spin-cavity observables. Here NGS uses $N_p = 12$, TWA is with $n_{\rm traj} = 10^4$ with standard error shaded.}
    \label{fig:Data_Ns3_Region_g=0.1_λ=1.0.pdf}
\end{figure*}

\begin{figure*}
    \centering
    \includegraphics[width=0.9\linewidth]{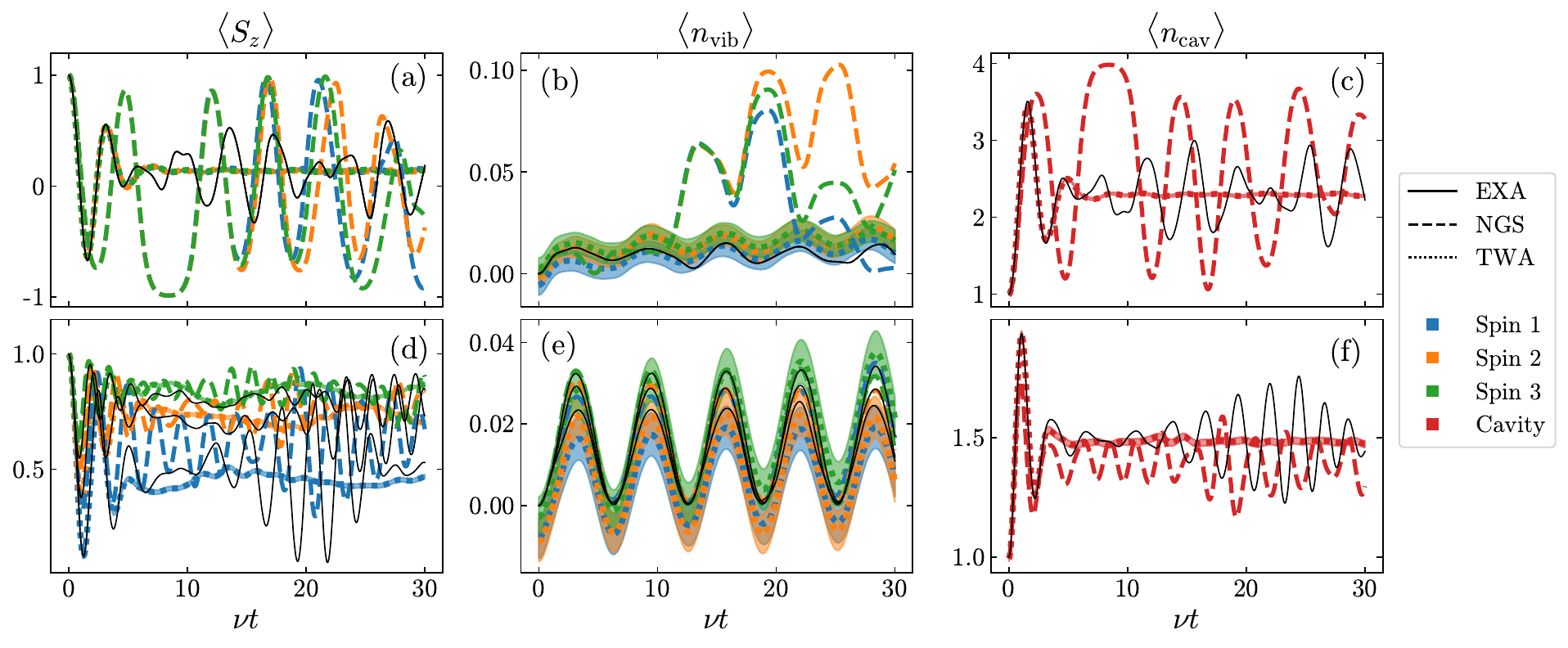}
    \caption{\emph{Closed dynamics}: $g = 1$, $\lambda = 0.1$, $\Delta = 0$. \emph{Top row}: without disorder. Beyond initial spin relaxation, both TWA and NGS perform relatively poorly. TWA incorrectly predicts equilibration of the spin-cavity dynamics. NGS does continue to produce spin-cavity dynamics, but overestimates the magnitude of the oscillations. \emph{Bottom row}: disordered, $\vec{\epsilon} = [2.6g,3.2g,4.2g]$. TWA incorrectly predicts equilibration of both spin and cavity observables after the first oscillation, whilst NGS produces qualitatively correct dynamics, even with only $N_p = 4$ coherent states. Here NGS uses $N_p = 2$ for the first row and $N_p = 4$ for the second row, and TWA is with $n_{\rm traj} = 10^4$ with standard error shaded.}
    \label{fig:Data_Ns3_Region_g=1.0_λ=0.1.pdf}
\end{figure*}
\begin{figure*}
    \centering
    \includegraphics[width=0.9\linewidth]{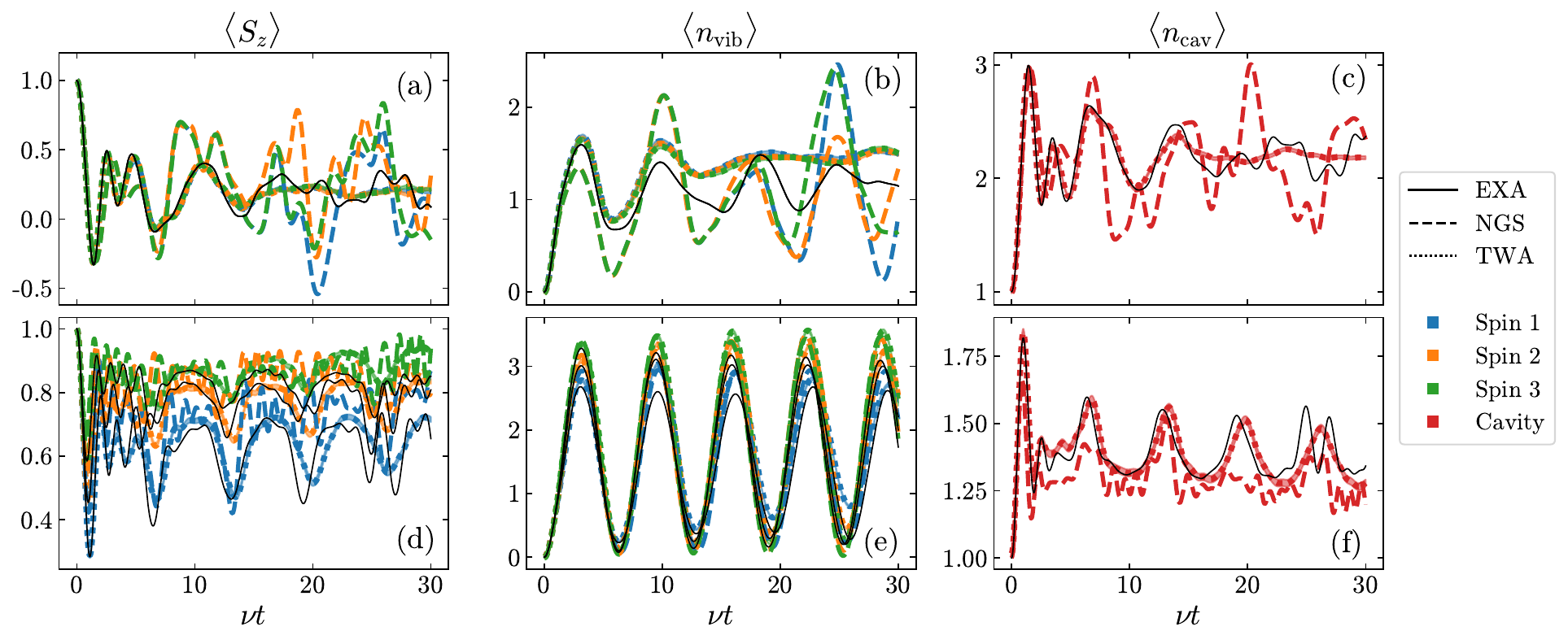}
    \caption{\emph{Closed dynamics}: $g = 1$, $\lambda = 1$, $\Delta = 0$. \emph{Top row}: without disorder. TWA accurately captures the dynamics at early times as compared to the NGS. 
    \emph{Bottom row}: disordered, $\vec{\epsilon} = [2.6g,3.2g,4.2g]$. TWA correctly captures key oscillations, even at late times. NGS also captures some of these features, but less accurately. Both methods perform similarly for the vibrational dynamics. 
    Here NGS uses $N_p = 4$, TWA is with $n_{\rm traj} = 10^4$ with standard error shaded.}
    \label{fig:Data_Ns3_Region_g=1.0_λ=1.0.pdf}
\end{figure*}

For all closed dynamics simulations, we set $\Delta = 0$. For evolution under the Holstein-Tavis-Cummings model in Eq.~(\ref{eq:H_molecule}), we find that while both methods are generally able to capture the short time dynamics, at later times they out-perform one another in different parameter regimes. We summarise our findings in Fig.~\ref{fig:schem}(a), where we qualitatively depict the performance for different parameter regimes in the absence of decoherence. More detailed dynamics for each regime are plotted in Figs.~\ref{fig:Data_Ns3_Region_g=0.1_λ=0.1.pdf}-\ref{fig:Data_Ns3_Region_g=1.0_λ=1.0.pdf}, with the first and second rows of each figure showing dynamics without and with disorder $\epsilon$, respectively. 

We begin in the weak coupling regime $g = \lambda = 0.1$, shown in Fig.~\ref{fig:Data_Ns3_Region_g=0.1_λ=0.1.pdf}. Here the dynamics is slow on the considered time-scale. In both the disorder-free (first row) and disordered (second row) settings, NGS and TWA accurately capture the small fluctuations of the vibrational modes at all times. Both methods also capture the initial spin relaxation, however NGS misses the revival time. TWA's ability to capture the first oscillation appears universally in all of the parameter regimes considered in this work. This behaviour can be understood as follows: due to our choice of a factorisable initial state with a corresponding positive semi-definite Wigner function, the TWA sampling is able to reproduce the initial state, and the mean-field and low order correlations that are generated during the short-time dynamics.

A generic feature of TWA is that when extending into the medium- to long-time dynamics, the potential buildup of higher order correlations is not captured by the method. While this not visible in Fig.~\ref{fig:Data_Ns3_Region_g=0.1_λ=0.1.pdf}, it can be seen clearly in the figures corresponding to the regimes discussed below.

The second regime we consider is the strong spin-vibrational coupling $\lambda \gg g$, shown in Fig.~\ref{fig:Data_Ns3_Region_g=0.1_λ=1.0.pdf}. Here we expect NGS to perform well, as NGS is exact with any coherent state number $N_p$ for $H_{\rm H}$. In the disorder-free regime (top row), although NGS with $N_p = 12$ does not fully capture the dynamics at late times, it does outperform TWA, which majorly underestimates the spin decay. In this case, the lack of disorder is challenging for our NGS ansatz: each spin evolves identically, requiring the superposition of coherent states to be factored into a product. This symmetry is broken when introducing disorder (second row), and we see that NGS captures accurately the dynamics for all considered times, including the initial decay and then revival of $\langle S_z \rangle$. For TWA, although the numerics match better for the vibrational dynamics in the presence of disorder, this is primarily a consequence of the fact that the disorder causes the Holstein interactions to dominate, and the spin and cavity dynamics continue to disagree with the exact solution.

Thirdly, we move to the strong spin-cavity coupling regime, $g \gg \lambda$, shown in Fig.~\ref{fig:Data_Ns3_Region_g=1.0_λ=0.1.pdf}. After accurately capturing the first oscillation, the TWA spin dynamics equilibrate about $\langle S_z \rangle \sim 0$ unlike the exact dynamics which, although they do oscillate about $\langle S_z \rangle \sim 0$, exhibit persistent oscillations with magnitude $\langle S_z \rangle \sim 1/2$. Similarly, NGS with $N_p = 4$ coherent states fails to capture any of the spin, vibration, or cavity dynamics. This is unsurprising because, in this regime where the Tavis-Cummings term dominates, we expect the number of coherent states required to scale as $\sim 3^{N_s}$, as each spin must be described using $N_p = 3$ coherent states. Although increasing $N_p$ would eventually improve the accuracy, we found that increasing it up to $N_p \lesssim 16$ did not provide a substantial increase in accuracy, whilst increasing the computational cost. In principle, TWA does not suffer from the same problem. However, if one needs to access higher order correlations with TWA, the introduction of higher order cumulants and the BBGKY hierarchy may become necessary. This poses an analogous problem to the number of Gaussian states in the ansatz: an exponentially increasing number of equations and potential numerical instabilities \footnote{In the numerical implementation of NGS we observe that for certain initial states the equations of motion become numerically unstable, which we attribute to the singularity of the metric $g$ and the associated need to perform a pseudoinverse instead of the inverse in the evaluation of $G = g^{-1}$. Dealing with this issue is a matter of current and future work}. 

\begin{figure*}
    \centering
    \includegraphics[width=0.9\linewidth]{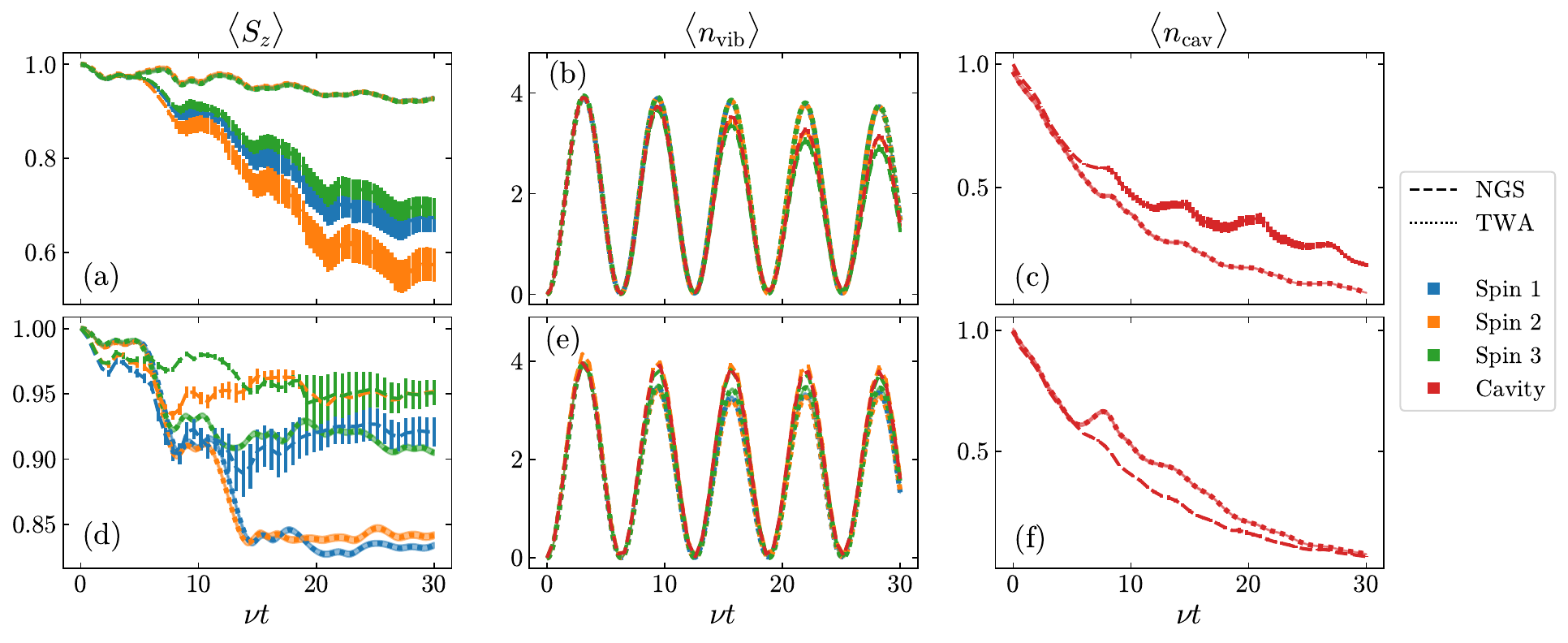}
    \caption{\emph{Open dynamics}. $g = 0.1$, $\lambda = 1$, $\Delta = 0$ and cavity decay $\kappa = g$. \emph{Top row:} without disorder. \emph{Bottom row:} with disorder $\vec{\epsilon} = [2g,3g,4g]$. NGS and TWA capture the same short-time dynamics, but disagree beyond $\nu t \sim 5$. Based on the corresponding closed dynamics results Fig.~\ref{fig:Data_Ns3_Region_g=0.1_λ=1.0.pdf}, we expect NGS to be more reliable in this regime. Exact numerics is challenging for open dynamics of systems of this size, see Fig.~\ref{fig:Ns1_OpenEDComparison} for benchmarking of smaller systems. Here NGS uses $N_p = 8$ and $n_{\rm traj} = 40$ with standard error indicated by the error bars, TWA is with $n_{\rm traj} = 10^4$ with standard error shaded.
    }
    \label{fig:decoweak}
\end{figure*}
\begin{figure*}
    \centering
    \includegraphics[width=0.9\linewidth]{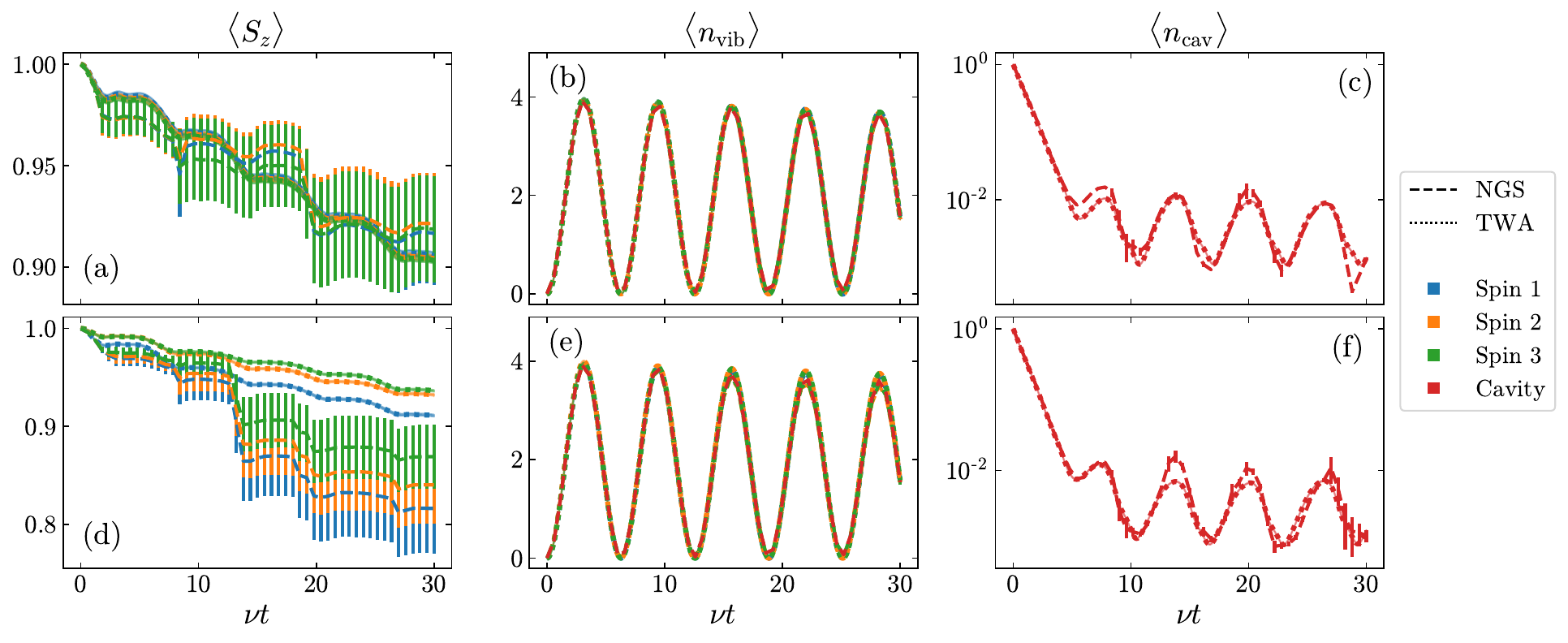}
    \caption{\emph{Open dynamics}. $g = 0.1$, $\lambda = 1$, $\Delta = 0$ and cavity decay $\kappa = 10g$. \emph{Top row:} without disorder. \emph{Bottom row:} with disorder $\vec{\epsilon} = [2g,3g,4g]$. We observe compatible results for the spins and a remarkable agreement between NGS and TWA for the vibrational and cavity dynamics, including the small amplitude oscillations in $\langle n_{\rm cav} \rangle$ at late times. Here NGS uses $N_p = 8$ and $n_{\rm traj} = 40$ with standard error indicated by the error bars, TWA is with $n_{\rm traj} = 10^4$ with standard error shaded.
    }
    \label{fig:decostrong}
\end{figure*}

Fourthly, we consider the strong spin-cavity and spin-vibration regime $\lambda = g = 1$, shown in Fig.~\ref{fig:Data_Ns3_Region_g=1.0_λ=1.0.pdf}. Without disorder, both methods struggle to capture the dynamics at late times, although TWA in particular is able to capture qualitative features with reasonable accuracy. Introducing disorder breaks the collective nature of the spins, enabling both methods to more accurately track the dynamics. TWA is able to qualitatively reproduce the periodic peaks in the spin and cavity dynamics at even later times than NGS. Both methods correctly obtain the vibrational dynamics.

Finally, we note that an advantage of NGS is the accessibility of the wavefunction. This means that any desired quantity, including entanglement entropy, can be computed. Furthermore, for small systems, strict performance measures such as the fidelity can be easily computed. These are shown in the supplementary material~\cite{supp} for the four different coupling regimes $(g,\lambda)$ considered in Figs.~\ref{fig:Data_Ns3_Region_g=0.1_λ=0.1.pdf}-\ref{fig:Data_Ns3_Region_g=1.0_λ=1.0.pdf}. The analogous plots for TWA cannot be generated.

\subsection{Open system dynamics}
\subsubsection{Holstein Tavis-Cummings}\label{sec:HolsteinTavisCummings}
Next, we introduce decoherence to our simulations. Figs.~\ref{fig:decoweak} and \ref{fig:decostrong} show the spin and bosonic dynamics for $N_s=3$ in the presence of cavity loss $a$ at strength $\kappa =g$ (Fig.~\ref{fig:decoweak}) and $\kappa =10 g$ (Fig.~\ref{fig:decostrong}), and in the regime $\lambda \gg g$, where NGS provides more accurate predictions in the closed setting (cf.~Fig.~\ref{fig:Data_Ns3_Region_g=0.1_λ=1.0.pdf}). In the top row we set the disorder $\epsilon=0$, while the bottom row shows the dynamics in the presence of disorder, $\vec{\epsilon} = [2g,3g,4g]$. Although exact dynamics were accessible for a closed system of this size, obtaining exact numerics in the open dynamics setting is challenging. In the supplementary material we compare the two methods against exact numerics for a smaller system of $N_s = 1$~\cite{supp}. 

For these parameters, we find that, perhaps unsurprisingly, NGS continues to perform well in both $\kappa=g$ and $\kappa=10g$ decoherence regimes. In the weaker decay limit shown in Fig.~\ref{fig:decoweak}, NGS and TWA agree only at short times. The under- and over-estimation of spin-cavity dynamics by TWA in the non-disordered and disordered systems respectively is consistent with the behaviour of TWA in the closed system, see Fig.~\ref{fig:Data_Ns3_Region_g=0.1_λ=1.0.pdf}. In the large decay limit shown in Fig.~\ref{fig:decostrong}, NGS and TWA are in reasonable agreement with one another for the spin dynamics and in near total agreement for the vibrational and cavity dynamics. Both show fast decay of the cavity to the vacuum, and remarkably both capture small oscillatory dynamics at late times with excellent agreement. Physically, both methods demonstrate that strong cavity decay stabilises the spin dynamics, which we attribute to the reduction of the effective Tavis-Cummings coupling strength and the prevention of the build up of correlations in the system between the spins, as well as spin-boson correlations, due to the loss of cavity excitations. 

\begin{figure}
    \centering
    \includegraphics[width=\linewidth]{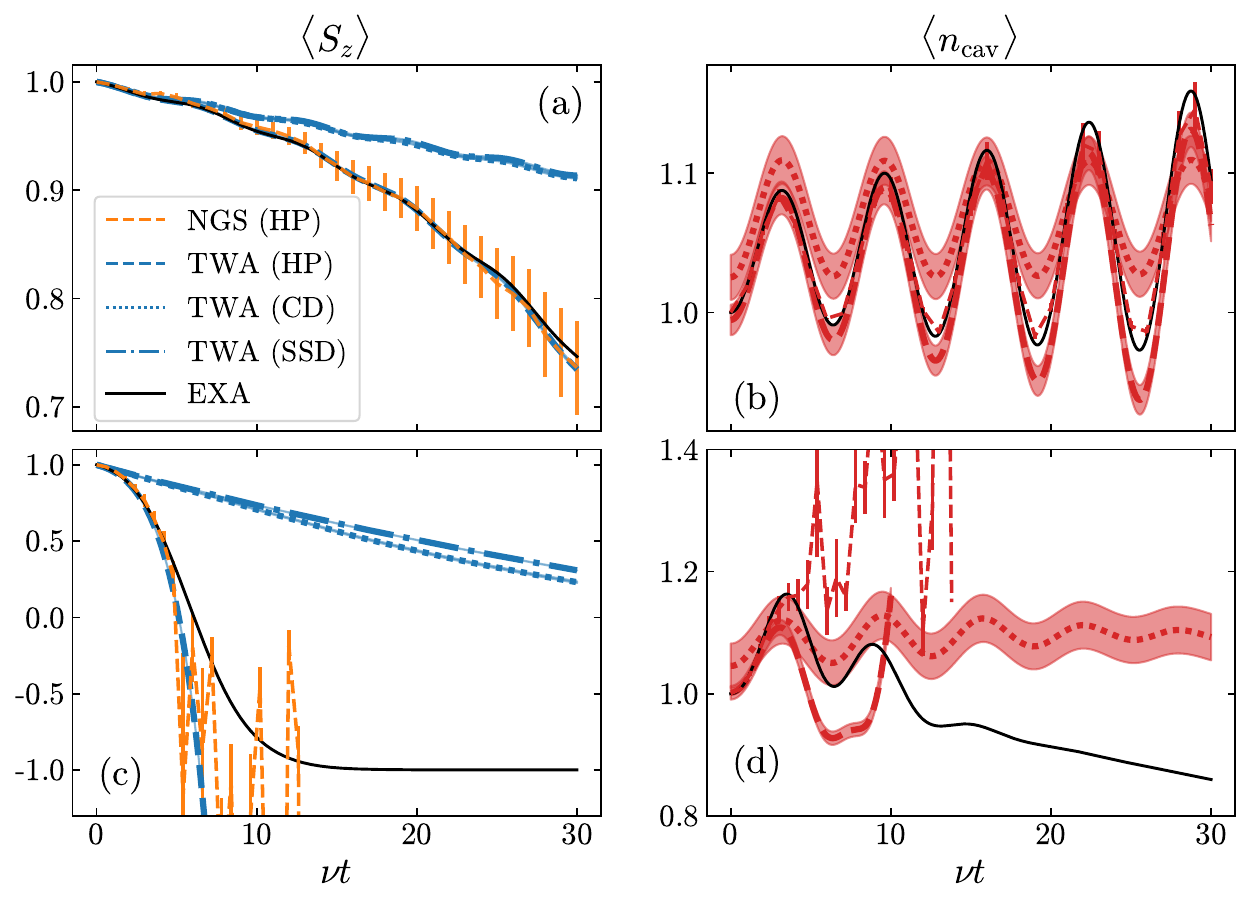}
    \caption{Collective spin dynamics. $\nu = \epsilon = \lambda = 0$, $\Delta = 1$, $g = 0.1$. \emph{Top row:} $\Gamma = 0.1g/\sqrt{N_s}$. NGS (HP) and TWA (HP) use the Holstein-Primakoff representation of the collective spin and agree excellently with exact numerics (EXA). On the other hand, when TWA treats the spins individually, the collective decay dynamics TWA (CD) agrees with the single spin decay dynamics TWA (SSD), but both deviate from the exact evolution due to TWA's failure to capture the correlations. \emph{Bottom row:} $\Gamma = g/\sqrt{N_s}$. NGS (HP) and TWA (HP) agree with exact numerics (EXA) until $t \nu \sim 5$, when the first order Taylor series expansion of the HP mapping breaks down. Here NGS uses $N_p = 8$ and $n_{\rm traj} = 40$ with standard error indicated by the error bars, TWA is with $n_{\rm traj} = 10^4$ with standard error shaded.}
    \label{fig:Data_Ns3_BigSpinDecay_g=0.1_Delta=1.0}
\end{figure}

\subsubsection{Tavis-Cummings}\label{sec:TavisCummings}
Next, we consider the effect of spin decay. We set $\epsilon, \, \lambda=0$ in the Hamiltonian~\eqref{eq:H_molecule} such that evolution is only under $H_{\rm TC}$, and include collective spin decay at rate $\Gamma$. We set $\Delta = 1$ and $g = 0.1$. Within the TWA formalism we treat the dynamics using two methods. First, we continue to treat the system as a collection of individual spins, as described in Sec.~\ref{sec:twa}. In Fig.~\ref{fig:Data_Ns3_BigSpinDecay_g=0.1_Delta=1.0} we plot the resulting dynamics where TWA (CD) refers to implementing this sampling in the presence of collective spin decay at $\Gamma=0.1g/\sqrt{N_s}$ and $g/\sqrt{N_s}$. TWA (SSD) uses the same sampling but the decay mechanism is single spin at the corresponding rate. The agreement between the two, and the disagreement with exact numerics (EXA) indicates that treating the spins individually with either of these two methods is inadequate to simulate collective spin decay. 

This motivates our second strategy. Because the Hamiltonian is $H_{\rm TC}$, the dynamics takes places in the collective Tavis-Cummings manifold as $H_{\rm TC}$ conserves the total excitation number $N_{\rm ex} = a^\dagger a + \sum_i \sigma_i^z$. Then, we can use the Holstein-Primakoff (HP) transformation to map the collective spin $S$ to a single bosonic mode, 
\begin{subequations}
\label{eq:HP_Usual}
\begin{align}
    S_+ &= \sqrt{2s-a_s^\dagger a_s}\, a_s, \\ 
    S_z &= s - a_s^\dagger a_s,
\end{align}
\end{subequations}
where $s \equiv N_s/2$. We use a Taylor series to expand the square root in powers of $1/s$ to first order. The NGS simulation then proceeds using the NGS ansatz with $N_b = 2$ bosonic modes. We can include collective spin decay $\sum_i \sigma_i^-$ at strength $\Gamma$ using the manifold projection technique described in Sec.~(\ref{sec:JumpsOutsideManifold}). 

For the TWA simulations, the equations of motion for the cavity mode $A$ and the large spin HP mode $B$ are, 
\begin{subequations}
\begin{align}
    \dot{A}_{\rm cl} &=- \frac{\kappa}{2} A_{\rm cl} - i\frac{g}{2\sqrt{2s}} B_{\rm cl}^*\left(4s-N_{B, \rm cl}\right),\\
    \dot{N}_{A, \rm cl} &=- \kappa N_{A,\rm cl} -\frac{g}{\sqrt{2s}} \Im[A_{\rm cl} B_{\rm cl}] \left( 4s-N_{B,\rm cl}\right),\\
    \begin{split}
    \dot{B}_{\rm cl} &=  \Gamma B_{\rm cl} \left[s-\frac{1}{8s} + N_{B, \rm cl} \left(\frac{\abs{B_{\rm cl}}^2 }{16s}-\frac{1}{2}\right)\right] \\
   & \hphantom{={}} + \frac{ig}{2\sqrt{2s}} \left[ A_{\rm cl} B_{\rm cl}^2 - 2 A_{\rm cl}^* \left(2s-N_{B, \rm cl} \right)  \right] +i \Delta B_{\rm cl},
    \end{split} \\
   \dot{N}_{B,\rm cl} &= \Gamma \left[2S+ N_{B, \rm cl} \left(\frac{1}{4s}+2\left(s-1 \right)+\mathcal{F}(B_{\rm cl})\right)\right]\nonumber\\
   &-\frac{g}{\sqrt{2s}}  \Im[A_{\rm cl}B_{\rm cl}] (4s-N_{B,\rm cl}),
\end{align}
\end{subequations}
where we introduce the function $\mathcal{F} (B_{\rm cl})=\frac{1}{8s}  \abs{B_{\rm cl}}^2\left(4-8s + \abs{B_{\rm cl}^2} \right)$.
We sample the cavity (mode A) assuming a coherent state $\ket{\alpha = 1}$ and sample the large spin (mode B) assuming the spins are polarized pointing up, which corresponds to sampling the vacuum for mode B. 

Our results using this approach, including collective spin decay, are shown in Fig.~\ref{fig:Data_Ns3_BigSpinDecay_g=0.1_Delta=1.0} in the lines labeled as TWA (HP) and NGS (HP). In the weak spin decay limit $\Gamma = 0.1g/\sqrt{N_s}$, both NGS (HP) and TWA (HP) are in excellent agreement with the exact numerics (EXA). NGS in particular captures the cavity dynamics with little error, whilst the error bars on the spin dynamics are still somewhat large due to our use of relatively few trajectories, $n_{\rm traj} = 40$. In the large decay limit $\Gamma = g/\sqrt{N_s}$, both NGS and TWA correctly capture the rapid spin decay until $\nu t \sim 5$. Beyond this point, the first order Taylor series expansion of the HP mapping breaks down, as highly excited Fock states are populated. One can potentially circumvent this issue by simulating collective spin decay as was performed in Ref.~\cite{Young2024} with DTWA. There, the spins were collectively coupled to a single cavity whose cavity loss was much stronger than the collective spin-cavity coupling, resulting in effective collective spin decay and without the utilization of the HP mapping. Comparing the performance of these approaches in different parameter regimes and for different models, e.g., for more complex forms of collective spin decay, represents an interesting future direction.

\section{Conclusions and outlook}\label{sec:conclusions}

\noindent\emph{Summary and conclusions:} In this work, we presented a non-Gaussian variational ansatz approach to studying the dynamics of open quantum systems composed of spin and bosonic degrees of freedom. While several other works in recent years have utilized NGS to study the time evolution of open quantum systems, previous efforts have focused on developing an equivalent ansatz for the density matrix and simulating the Lindblad equations. Here, we utilized the quantum trajectories method, allowing us to take advantage of the previously developed machinery and analytic expressions obtained for real- and imaginary-time dynamics. 

In addition to providing a comprehensive overview of this method, we performed extensive numerical simulations over a broad range of parameters of a spin-boson Hamiltonian [Eq.~\eqref{eq:H_molecule}] with Tavis-Cummings (TC) and Holstein couplings, which is applicable to a broad range of quantum simulation platforms as well as problems of interest in quantum chemistry, atomic physics and condensed matter theory. We compared the performance of NGS with a method using the truncated Wigner approximation for systems with mixed bosonic and spin degrees of freedom, extended to open quantum systems following the approach in Ref.~\cite{FleischhauerDTWA}. 

In the absence of decoherence our findings are as follows: for strong TC coupling, TWA is the more accurate method, while for strong Holstein couplings NGS is the better choice. When neither term dominates, for both weak and strong coupling regimes, TWA captures the short-time dynamics, while NGS generally displays the correct qualitative behaviour, even at late times. After introducing spin disorder the performance of NGS typically improves, whilst for TWA the vibrational dynamics match the exact dynamics better, which is attributable to the fact that disorder causes the Holstein interaction to dominate. 

For open quantum dynamics we focused on the regime where the Holstein term dominates and considered the effect of cavity loss. At weak decay rates the NGS continues to perform well. TWA improves as the cavity decay rate increases due to the loss of quantum correlations, with both NGS and TWA showing excellent agreement. In the presence of collective spin decay we considered the TC model only, finding that in the limit of small collective decay, both methods perform well when using a Holstein-Primakoff transformation for the large spin. In the limit of large collective decay, NGS and TWA both only capture the short time dynamics as the Holstein-Primakoff transformation is no longer accurate at later times. Using TWA we were able to also treat each spin individually. However, TWA does not capture the collective nature of the decay, with the results closely matching the effect of single spin decay. 

Further considerations should also be made when deciding between the two methods. Although the NGS ansatz can be made less computationally demanding by reducing the number $N_p$ of coherent states, in general TWA methods are easier to implement and require fewer computational resources. The resource requirement and the complexity of NGS is offset by the advantages that it is a controlled approximation and gives access to the wavefunction, allowing one to access any observable, including higher-order correlations. The TWA framework needs to be amended if one hopes to capture these correlations accurately. A potential strategy to remedy this can be to use a cluster TWA approach~\cite{wurtz2018cluster}. There, several sub-system parts are grouped into a single (discrete or continuous) large sub-system that, provided a proper sampling strategy, follows the fully exact quantum evolution, while correlations between clusters are approximated in TWA. Such approaches allow to use the cluster size as controllable parameter to enhance the simulation towards the exact one.

\noindent\emph{Outlook:} The extension of NGS to open quantum systems using the quantum trajectories formulation that we presented in this work is perhaps the most natural pathway. For Hermitian jump operators, an alternative approach and a simple extension of this work would be to instead solve the stochastic partial differential equations that result from the unravelling of the master equation. Furthermore, one could explore the impact of the chosen unraveling on the performance of the NGS method at a fixed $N_p$, as it has been shown that this choice can have a large effect on the entanglement buildup in the trajectory~\cite{Vovk2022, vovk2024quantumtrajectoryentanglementvarious}. 

Another limitation of the present formulation is that the spin states in Eq.~(\ref{eq:psi}) are exact and span the full spin Hilbert space of dimension $2^{N_s}$, thus limiting the use of the ansatz to a handful of spins unless approximations such as the large $N_s$ expansion in the Holstein-Primakoff mapping used in Sec.~\ref{sec:TavisCummings} are invoked. On the one hand, studies using a fermionic Gaussian state representation of spins have been performed \cite{Kaicher_2023_PRB}, and these might be combined in principle in a straightforward way with the NGS ansatz for bosons \cite{Shi_2018_AnnPhys}. On the other hand, it would be highly interesting to combine the NGS ansatz with other variational techniques highly suitable for the spins such as tensor network based approaches \cite{Schollwock_2011_AnnPhys, Orus_2014_AnnPhys, Montangero_2018_Book}. Furthermore, similar to the present comparison between NGS and TWA, it would be beneficial to apply the here-presented non-Gaussian ansatz to the study of other systems which might be challenging to simulate otherwise. These include for instance purely bosonic models, such as the Bose-Hubbard model, with disorder and on non-regular lattices \cite{Gottlob_2023_PRB}. Such extensive studies will allow for a comparison between our approach and the corresponding master equation approach based on extending the ansatz of Eq.~(\ref{eq:psi}) to density matrices~\cite{Schlegel_2023, Joubert_2015_JChemPhys}. 

Finally, we note that a particular promising application field of the methodologies introduced here could be in the emerging field of polaritonic chemistry~\cite{hutchison2012modifying, simpkins2021mode, wang2021a, campos2023swinging}. There, recent experiments have demonstrated that large {\em collective} strong cavity couplings (e.g.~$g_c = g \sqrt{N}$) can be functionalized to modify chemical reactivity. A theoretical understanding for such modifications are currently centered around the question how delocalized polaritonic state can play a role for  changing chemistry on the single-molecule level, as local amplitudes of collective polaritons vanish in the thermodynamic limit. In spin-boson approximations to the problems, in particular for the disordered Holstein-Tavis-Cummings~\cite{herrera2016cavity} model that we studied here, it was recently discovered that the interplay of disorder and collective cavity couplings can give rise to robust local quantum effects in the large-$N$ limit in the form of non-Gaussian distributions of the nuclear coordinate~\cite{Wellnitz_2022CommPhys}. Using matrix product state methods, it was possible to push simulations to systems with 160 effective molecules, but in particular the TWA approach discussed here would allow access to much larger systems. Typical parameter regimes discussed here are well covered by the TWA approach (e.g.~the typical parameters from~\cite{Wellnitz_2022CommPhys} corresponds to $\lambda \sim 0.1 \nu$ and $g \ll \nu$) and thus hint to a general applicability of the method in the relevant regime. Further, the TWA approach can be straightforwardly adapted to simulate nuclear dynamics not only on harmonic but also arbitrary potential energy surfaces, whilst the NGS ansatz using the machinery presented here can also model anharmonic Hamiltonian terms. In the future this may allow for the analysis of quantum effects in realistic chemical reaction models, even in a macroscopic limit.

\section{Supplementary Material}
In the supplementary material, we show that the tangent space of the variational manifold for the NGS ansatz is a K\"{a}hler manifold, and benchmark NGS and TWA against exact numerics for a single spin $N_s = 1$~\cite{supp}. 

\begin{acknowledgments}
A.S.N., B.G., L.J.B., and J.M. are supported by the Dutch Research Council (NWO/OCW), as part of the Quantum Software Consortium programme (project number 024.003.037), Quantum Delta NL (project number NGF.1582.22.030), and ENW-XL grant (project number OCENW.XL21.XL21.122). J.M. was partly supported by a Proof of Concept grant through the Innovation Exchange Amsterdam (IXA) and the NWO Take-off Phase 1 grant (project number 20593). 
J.T.Y. is supported by the NWO Talent
Programme (project number VI.Veni.222.312), which is
(partly) financed by the Dutch Research Council (NWO). 
J.S. is supported by the Interdisciplinary Thematic Institute QMat, as part of the ITI 2021-2028 program of the University of Strasbourg, CNRS and Inserm, and was supported by IdEx Unistra (ANR-10-IDEX-0002), SFRI STRAT'US project (ANR-20-SFRI-0012), and EUR QMAT ANR-17-EURE-0024 under the framework of the French Investments for the Future Program. Work was supported by the CNRS through the EMERGENCE@INC2024 project DINOPARC and by the French National Research Agency under the Investments of the Future Program project ANR-21-ESRE-0032 (aQCess).
\end{acknowledgments}

\bibliography{bib,DTWA}

%%%%% SUPPLEMENTARY %%%%%

\clearpage

\setcounter{section}{0}
\setcounter{equation}{0}
\setcounter{figure}{0}
\setcounter{table}{0}
\setcounter{page}{1}

% \counterwithin{figure}{section}
\renewcommand\thefigure{\arabic{figure}}

\let\theequationWithoutS\theequation % <- store old definition
\renewcommand\theequation{S\theequationWithoutS}
\let\thefigureWithoutS\thefigure % <- store old definition
\renewcommand\thefigure{S\thefigureWithoutS}
\let\thesectionWithoutS\thesection % <- store old definition
\renewcommand\thesection{S\thesectionWithoutS}

\title{Supplementary Material: \mytitle}
\maketitle
\onecolumngrid

\section{On the complex structure of the NGS ansatz}
\label{sec:Complex_structure}
In this section we specify the geometric structures for special cases of the general NGS ansatz. We begin with the example of a coherent state ansatz, i.e. $N_p = 1$, with explicit normalization and phase factors, $\ket{\psi(\vec{z})} = e^{\kappa + i \theta}\mathcal{D}(x+iy)\ket{0}$, with $\vec{z} = \{\kappa,\theta,x,y\}$. The tangent vectors for this ansatz were given in Eq.~\eqref{eq:CohStateAnsatzTangentVectors}, from which we can compute the metric and symplectic forms, 
\begin{align}
    \qquad g_{\mu \nu} = \left(
    \begin{array}{cccc}
        1 & 0 & 0 & 0 \\
        0 & 1 & -z_4 & z_3 \\ 
        0 & -z_4 & 1 + z_4^2 & -z_3 z_4 \\ 
        0 & z_3 & -z_3 z_4 & 1 + z_3^2
    \end{array}
    \right), \qquad \omega_{\mu \nu} = \left( 
    \begin{array}{cccc}
       0  & 1 & - z_4 & z_3 \\
       -1 & 0 & 0 & 0 \\ 
       z_4 & 0 & 0 & 0 \\ 
       -z_3 & 0 & -1 & 0
    \end{array}
    \right)
\end{align}
as well as their respective inverses
\begin{align}
    \qquad G = \left(
    \begin{array}{cccc}
        1 & 0 & 0 & 0 \\
        0 & 1 + z_3^2 + z_4^2 & z_4 & -z_3 \\ 
        0 & z_4 & 1 & 0 \\ 
        0 & -z_3 & 0 & 1
    \end{array}
    \right), \qquad \Omega = \left( 
    \begin{array}{cccc}
       0  & -1 & 0 & 0 \\
       1 & 0 & -z_3 & -z_4 \\ 
       0 & z_3 & 0 & -1 \\ 
       0 & z_4 & 1 & 0
    \end{array}
    \right).
\end{align}
Note that here the use of the pseudo-inverse would give the same as the inverse, as $\omega_{\mu \nu}$ and $g_{\mu \nu}$ are not singular for any values of the variational parameters $\vec{z}$. We can also compute the complex structure, $J\indices{^\mu_\nu} = - G^{\mu \sigma} \omega_{\sigma \nu}$, 
\begin{align}
J\indices{^\mu_\nu} = \left(
    \begin{array}{cccc}
        0 & -1 & z_4 & -z_3 \\
        1 & 0 & -z_3 & z_4 \\ 
        0 & 0 & 0 & -1 \\ 
        0 & 0 & 1 & 0
    \end{array}
    \right),
\end{align}
and verify that $J^2 = -\mathbb{1}$. Using the relation $i \ket{v_\nu} = J\indices{^\mu_\nu} \ket{v_\mu}$ which holds if the tangent space of the variational manifold is a K\"{a}hler manifold as is the case here, we directly obtain the relations 
\begin{align}
    i \ket{v_1} &= \ket{v_2}, \\ 
    i \ket{v_2} &= -i \ket{v_1}, \\ 
    i \ket{v_3} &= z_4 \ket{v_1} - z_3 \ket{v_2} + \ket{v_4}, \\ 
    i \ket{v_4} &= -z_3 \ket{v_1} + z_4 \ket{v_2} - \ket{v_3}, 
\end{align}
which we derived explicitly and used in the main text to compute $\text{Im}[C_\mu]$, see Sec.~\ref{sec:Open_EoMs}.

Next, we compute the complex structure of the squeezed coherent state, including norm and phase factors, $\ket{\psi(\vec{z})} = {\rm e}^{\kappa+i \theta} \ket{\alpha,\zeta} = {\rm e}^{\kappa+i \theta} D(\alpha) \xi(\zeta) \ket{0}$, such that $\vec{z} = (\kappa, \theta, x,y,r,\phi)$ with $\alpha = x+i y$, $\zeta = r {\rm e}^{i \phi}$. Using the expressions in Eqs.~(\ref{eq:dxUa}), (\ref{eq:fgh}) we arrive at
\begin{align}
   J = \left(
\begin{array}{cccccc}
 0 & -1 & y & -x & 0 & -\sinh ^2(r)/2 \\
 1 & 0 & t_1 & t_2 & -\tanh (r)/2 & 0 \\
 0 & 0 & \sinh (2 r) \sin (\phi) & - \cosh(2 r) - 2\cos(\phi)\cosh(r)\sinh(r) & 0 & 0 \\
 0 & 0 & \cosh (2 r) - 2 \cos (\phi) \cosh(r) \sinh(r) & -\sinh(2 r) \sin(\phi) & 0 & 0 \\
 0 & 0 & 0 & 0 & 0 & -\sinh (r) \cosh (r) \\
 0 & 0 & 0 & 0 & \text{csch}(r) \sech(r) & 0 \\
\end{array}
\right),
\label{eq:J_squeezed}
\end{align}
where we introduced $t_1 = -x \cosh(2r) + (x \cos(\phi) + y \sin(\phi)) \sinh(2r) $ and $t_2 = -y \cosh(2r) + (x \sin(\phi) - y \cos(\phi))\sinh (2 r) $. It is straightforward to verify that
\begin{align}
    J^2 = -1, 
\end{align}
i.e. the tangent space of the single-mode, single-Gaussian state ansatz is a K\"{a}hler space. A technical remark is in order. We note that 
\begin{align}
    \det(g) = \det(\omega) \propto \sinh(2r)^2, \label{eq:DetSameNp}
\end{align}
which vanishes in the limit of vanishing squeezing $r \rightarrow 0$ and consequently $G = g^{-1}$ is ill-defined (an alternative way of seeing this is an overparametrization of the tangent vector space as $\ket{v_6}$ vanishes, cf. the Eqs.~(\ref{eq:v_squeezed_coherent}) below). When such situation occurs, one can use instead the Moore-Penrose pseudo-inverse, as suggested in \cite{Hackl_2020_SciPost}, to evaluate $g^{-1}$. In this case we recover $J^2 = -1$ and thus the K\"{a}hler space. 
Finally, when using Eq.~(\ref{eq:J_squeezed}) in Eq.~(\ref{eq:projection}) we arrive at the following tangent vectors (in addition to $\{ \ket{v_\mu} \} $)
\begin{subequations}\begin{align}
    i \ket{v_1} &= \ket{v_2}, \\ 
    i \ket{v_2} &= - \ket{v_1}, \\ 
        i \ket{v_3} &= y \ket{v_1} + t_1 \ket{v_2} + \sinh(2r)\sin(\phi) \ket{v_3} + [ \cosh(2r) - 2\cos(\phi) \cosh(r)\sinh(r) ] \ket{v_4}, \\ 
    i \ket{v_4} &= -x \ket{v_1}  + t_2 \ket{v_2} - [\cosh(2r) + 2 \cos(\phi)\cosh(r)\sinh(r)] \ket{v_3} - \sinh(2r)\sin(\phi)\ket{v_4}, \\ 
    i \ket{v_5} &=  - (1/2) \tanh(r) \ket{v_2} + \text{csch}(r) \sech(r)\ket{v_6}, \\ 
    i \ket{v_6} &= - (1/2)\sinh^2(r) \ket{v_1} - \sinh(r) \cosh(r) \ket{v_5}.
\end{align}\label{eq:v_squeezed_coherent}\end{subequations}
It is apparent from the above equations that the set $\{i\ket{v_\mu}\}$ lies in the tangent space spanned by $\{ \ket{v_\mu} \}$ as a consequence of it being the K\"{a}hler manifold, i.e. $J^2=-1$.

The same procedure then generalizes to the construction of tangent vectors for each Gaussian $p \in \{1,\ldots,N_p\}$ and each mode $k \in \{1,\ldots,N_b\}$ such that the tangent space remains K\"{a}hler manifold also in the generic NGS ansatz of a superposition of many-mode squeezed displaced states. In this case the complex structure $J$ takes a block-diagonal form, where each block is labeled by the Gaussian and mode number $(p,k)$ with the property $J^2=-1$, which we could also verify numerically.

\section{Benchmarking dynamics with single spin}
\begin{figure}
    \centering
    \includegraphics[width=0.44\linewidth]{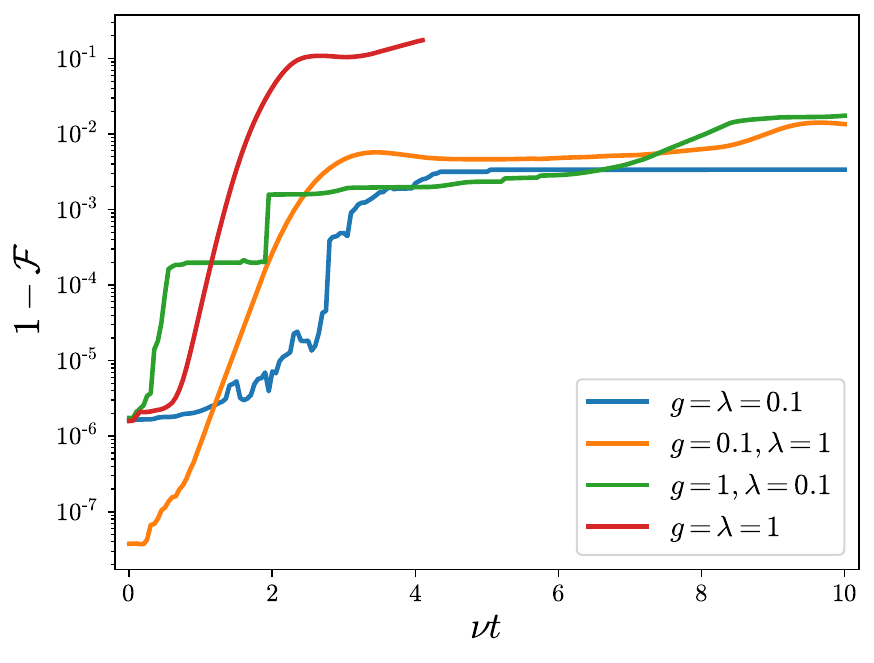}
    \caption{Infidelity $1 -\mathcal{F}(t) = 1 - |\braket{\psi(t)}{\Psi(t)}|^2$ of real-time dynamics for $N_s = 1$ between NGS $\ket{\psi}$ using $N_p = 8$ and numerically exact $\ket{\Psi}$ for the same four parameter regimes studied in the main text. The initial infidelity is due to introducing a tiny randomness in the NGS initial state of the bosonic modes, which is a strategy we adopt to lift the degeneracy between the different multi-mode coherent states to avoid overparametrization of the equations of motion, cf. \cite{Bond_2024_PRL}. In all regions except $g = \lambda = 1$ where the rapid decrease in infidelity leads to unstable NGS EOMs, NGS captures the dynamics even at late times with excellent infidelity, $1-\mathcal{F} \lesssim 10^{-2}$.}
    \label{fig:Ns1_FidelityPlot}
\end{figure}

A key advantage of NGS is that it provides access to the wavefunction $\ket{\psi}$, meaning that any desired quantity can be computed, including for example the entanglement entropy. We provide a simple example of this utility by computing the infidelity $1-\mathcal{F}(t) = 1 - |\braket{\psi(t)}{\Psi(t)}|^2$ between the NGS wavefunction $\ket{\psi}$ with $N_p = 8$ and the exact wavefunction $\ket{\Psi}$ for closed real-time dynamics. This is shown in Fig.~\ref{fig:Ns1_FidelityPlot} for $N_s = 1$ for the four parameter regimes studied in Figs.~\ref{fig:Data_Ns3_Region_g=0.1_λ=0.1.pdf}-\ref{fig:Data_Ns3_Region_g=1.0_λ=1.0.pdf} (there with $N_s = 3$). Note that such a comparison for more spins becomes challenging as we cannot access exact numerics. We find that in all regimes except $g = \lambda = 1$, the infidelity of NGS is low, with $1 - \mathcal{F} < \mathcal{O}(10^{-2})$. We also note that the infidelity is not accessible in TWA. 

\begin{figure}
    \centering
    \includegraphics[width=\linewidth]{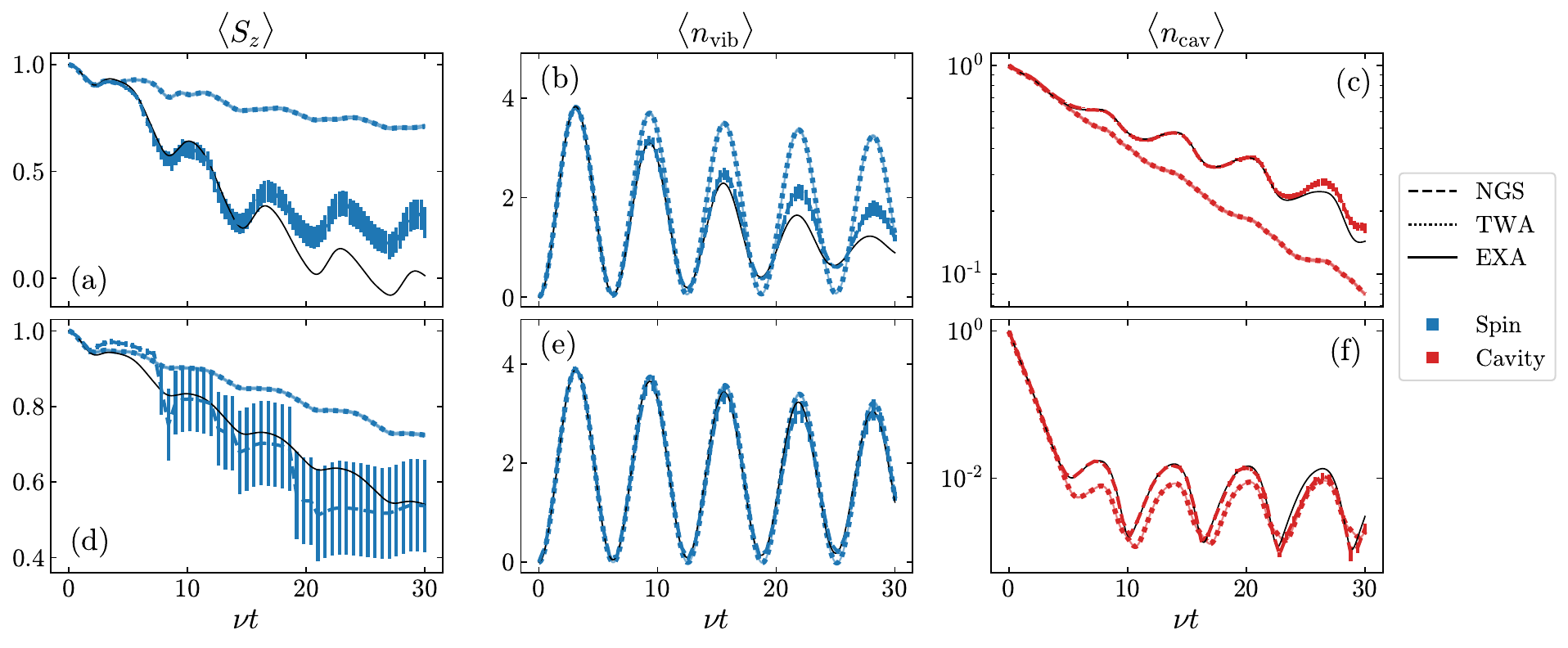}
    \caption{Benchmarking NGS and TWA against exact numerics. We use $N_s = 1$ spin in the $g = 0.1$, $\lambda = 1$ regime with cavity loss $\sqrt{\kappa} a$, i.e. the same parameters considered in the main text (there with $N_s = 3$). \emph{Top row:} weaker cavity decay $\kappa = g$. NGS accurately captures the cavity dynamics, and captures the spin and vibrational mode decay until $\nu t \sim 15$. TWA significantly under-estimates the spin and vibrational decay, while over-estimating the cavity decay. \emph{Bottom row:} strong cavity decay $\kappa = 10g$. Both methods capture the decay of cavity population and subsequent small oscillations, as well as the vibrational loss. NGS now agrees with exact numerics for the spin decay (albeit with large error bars due to the limited $n_{\rm traj})$. TWA still underestimates this observable, although the agreement is closer. We attribute this to the loss of quantum correlations due to the decoherence which makes it easier for the TWA to capture the true quantum state. Here NGS uses $N_p = 8$ and $n_{\rm traj} = 40$ with standard error indicated by the error bars, TWA is with $n_{\rm traj} = 10^4$ with standard error shaded.}
    \label{fig:Ns1_OpenEDComparison}
\end{figure}

Next, we benchmark both TWA and NGS against exact numerics for open real-time dynamics. In Fig.~\ref{fig:Ns1_OpenEDComparison}, we consider $N_s = 1$ in the same parameter regime as the main text $g = 0.1, \lambda = 1$ with cavity loss $\sqrt{\kappa} a $ at rates $\kappa = g$ (top row) and $\kappa = 10g$ (bottom row). For weaker cavity loss $\kappa = g$ we find that NGS with $N_p = 8$ coherent states more closely tracks the exact numerics than TWA, particularly for the cavity dynamics which shows excellent agreement. At strong cavity decay rates $\kappa = 10g$ both methods accurately capture the cavity dynamics, with NGS also capturing the spin decay, albeit with large error bars as only $n_{\rm traj} = 40$ trajectories were used. 

\end{document}